\documentclass[amsmath,amssymb,aps,reprint,prb]{revtex4-2}
\usepackage{subcaption}
\usepackage{rotating,booktabs}
\usepackage{graphicx}% Include figure files
\usepackage{bm}% bold math
\usepackage{tabularx}
\usepackage{braket}
\usepackage{amsfonts}
\newcommand{\shortminus}{ - }

\usepackage{hyperref}
\hypersetup{
    colorlinks=true,
    linkcolor=blue,
    citecolor=blue,
    filecolor=blue,      
    urlcolor=blue,
    pdfpagemode=FullScreen,
    }
\begin{document}

%\preprint{APS/123-QED}

\title{Empirical Tight-Binding Parameters for Wurtzite  group III-V(non-Nitride) and IV Materials}% 

\author{Joseph Sink}
 \email{joseph-sink@uiowa.edu}
 \affiliation{ Department of Physics and Astronomy, University of Iowa}

\author{Craig Pryor}
\affiliation{ Department of Physics and Astronomy, and Optical Science and Technology Center,  University of Iowa}
 \email{craig-pryor@uiowa.edu}

\date{\today}

\begin{abstract}
Many commonly used nearest neighbor tight binding models for cubic semiconductors often result in inaccurate band structures when transferred to hexagonal polytypes. 
The resulting bandgaps are systematically too small, and in some cases calculations may erroneously predict a metal for the wurtzite polytype.
We have calculated a set of $\rm spds^*$ tight binding parameters for the hexagonal phases of non-nitride III-V and group V semiconductors fit to experimental data and empirical pseudopotential calculations. 
Our fitting procedure constrains the parameters to be as close as possible to those for the cubic polytype so as to make the the parameters maximally transferable.
Our parameters, when combined with the existing cubic parameters, provide a model suitable for modeling the electronic structure of polytypic heterostructures containing both the wurtzite and zincblende crystal phases.
\end{abstract}
\maketitle
\section{Introduction}

Non-nitride III-V and group V semiconductor materials are well known to have cubic symmetry (3C-zincblende or diamond) when grown in bulk, but may exhibit exotic crystal phases (2H-wurtzite or lonsdaleite, 4H, etc) when grown in high aspect ratio morphologies such as nanowires. By carefully tuning growth conditions\cite{Dubrovskii.NanoLetters.21,Yuan.APR.21,Wang.Small.21,Staudinger.NanoLett.20,Martensson.CrystGD.20}, it is possible to deliberately alternate between the cubic and hexagonal phases to form polytypic heterostructures in which the chemical composition is constant throughout the structure. While chemically identical, the differences in the bonding geometry between polytypes gives rise to differences in their electronic properties. There has been a great deal of interest in this new and novel degree of freedom.
\cite{Assali.NanoLatters.13,Perera.NanoLetters.13,Kriegner.IOPNanoTech.11,DeLuca.NanoLetters.17,Dubrovskii.NanoLetters.21,Yuan.APR.21,Wang.Small.21,Jacobsson.CGD.15,Staudinger.NanoLett.20,Martensson.CrystGD.20,Pournia.NanoRes.2020,Caroff.NatNano.09,Amato.NanoLetters.16,Fasolato.NanoLetters.18,Chen.NanoLetters.17,Ketterer.ACSNano.11}

Extracting bulk electronic properties from wuzrtite samples is complicated by their small dimensions, resulting in a lack of experimental data available to constrain empirical models. Photoluminescence, high-field magneto-photoluminescence spectroscopy\cite{DeLuca.NanoLetters.17} and polarized photocurrent spectroscopy\cite{Pournia.NanoRes.2020} can be used to measure the bandgap, band symmetry ordering and exciton effective mass as a function of position along the nanowire. However, many other properties currently rely on theoretical calculations based on highly transferable models such as density functional theory\cite{Heiss.PRB.11} and empirical pseudopotentials\cite{De.PRB.2010,De.IOP.2014}.
For small changes in the local electronic environment, pseudopotentials that are empirically fit to the band structure of the cubic phase may be transferred to the positions of the atoms in hexagonal phase without changing them.
The band structures resulting from this method are in good agreement with existing experimental results and provide additional band structure information\cite{De.PRB.2010,De.IOP.2014}.

While pseudopotentials have been used to directly compute the electronic structure of semiconductor nanostructures\cite{Wang.prb.1996, Wang.prb.1996b}, it has been more common to use the tight binding approach, in which the single electron wave function in a material is expressed as a sum of localized orbitals centered on the atoms.
In the Empirical Tight-Binding (ETB) approach 
\cite{Schulz.PRB.06,
Zielinski.prb.2010,  % TB QD
Scholz.Jancu.PSS.00,Cygorek.PRB.20} the Hamiltonian matrix elements between localized orbitals are treated as parameters that are fit to experimental or {\it ab initio} results for bulk materials.
For inhomogeneous systems the Hamiltonian matrix elements between localized orbitals are assumed to be the same as in the bulk, depending only on the atoms begin coupled.
Additional approximations are usually made for computational efficiency:
1) only nearest neighbor (NN) orbitals are coupled 2) Hamiltonian matrix elements involving orbitals on two different atoms and the potential from a third atom (three-center integrals) are ignored 3) the orbitals on different atoms are assumed to be orthogonal. The last of these may be ensured by Lo{\"w}din orthogonalization\cite{Aiken.IJQC.80} but the first two are true approximations.

Tight binding parameters are said to be transferable when the Hamiltonian matrix elements between orbitals can be transferred between similar materials.
This would mean, for example, that the matrix elements between orbitals in adjacent Ga and As atoms would be the same in GaAs and in an AlGaAs alloy.
This relationship is only approximate since the material surrounding the Ga-As pair will effect the local electronic environment, altering the local atomic wavefunctions and thereby the strength of the hopping interactions. The effects of this perturbation are generally small between like materials, and manifest as chemical trends such as those observed in Ref. \onlinecite{Jancu.prb.1998}.

The cubic and hexagonal phases of III-V (zincblende and wurtzite) and group IV (diamond and lonsdaleite) semiconductors would appear to be ideal candidates for polytype transferability.
Not only are the materials chemically identical, but the four nearest neighbors and 9 out of 12 second nearest neighbors are at the same positions, with very small deviations that can be treated with strain modifications. 
The similarities of the polytypes are confirmed by the small differences in experimental bandgaps between polytypes.
The agreement between pseudopotential calculations and experiment suggests that the crystal potentials of the two polytypes are very similar, with a rearrangement of the atomic positions leading to subtle changes in the band structure.

However, transferring parameters from the widely used cubic $spds^*$ model of Ref. \onlinecite{Jancu.prb.1998} to hexagonal crystals results in bulk band structures in significant disagreement with experiment\cite{Vainorius.Nanoscale.18,Pournia.NanoRes.2020,Chauvin.APL.18,Kriegner.PRB.13} and theory\cite{De.PRB.2010,De.IOP.2014}. 
%This indicates that the hexagonal polytypes (2H, 4H, 6H,etc) require different parameters which must be fit to reliable target data.
This indicates that the hexagonal polytypes require different parameters which must be fit to reliable target data. 
It has been observed for SiC using an $sps^*$ model that excellent transferability may be expected among the hexagonal 2H, 4H, 6H and 8H polytypes \cite{Laref.PSS.08}.
We have carried out such a procedure, fitting $spds^*$ tight binding parameters to experimental data when it is available and to empirical pseudopotential band structures when experimental data is not available.
We impose a penalty function in the fitting procedure to keep the parameters for the hexagonal phase as close to the cubic parameters as possible.
Our result is a set of tight binding parameters for the non-nitride III-V and group IV semiconductors in the hexagonal phase that is still relatively close to the parameters for the cubic phase, thus maintaining {\it approximate} transferability.

Section \ref{sec:background} reviews the relevant formalism and notation we use. 
Section \ref{Hamiltonian} discusses the ETB, the approximations typically made, and their implications for transferability. 
In section \ref{sec:methods} we describe our fitting procedure in some detail, and present the results. Section \ref{sec:ResultsAndDiscussion} concludes with a discussion of possible applications of our new parameters and what the lack of transferability implies for the accuracy of tight binding models in general.

\section{Background}\label{sec:background}

\subsection{Geometry: WZ and ZB}\label{Crystals}

The WZ crystal is defined by four interpenetrating hexagonal lattices, with conventional lattice vectors
\begin{align}
    &\vec{a}_1=\{\frac{1}{2},\frac{\sqrt{3}}{2} , \}a 
    &\vec{a}_2=\{\frac{1}{2},-\frac{\sqrt{3}}{2} , 0\}a \nonumber \\
    &\vec{a}_3=\{0,0,c\}
\end{align}
and basis vectors
\begin{align}
    &\vec{t}_1=\{0,0,0\}                     &&   \vec{t}_2=\{0,0, uc\}\nonumber\\
    &\vec{t}_3=\{0,\frac{\sqrt{3}a}{3},\frac{c}{2}\} &&   \vec{t}_4=\{0,\frac{\sqrt{3}a}{3},(\frac{1}{2}+u)c\}
\end{align}
where $a$, $c$ and $u$ are material dependent lattice constants. In the ideal WZ configuration, all nearest neighbor bond lengths ($|\vec{t}_1 - \vec{t}_2|=|\vec{t}_2\shortminus\vec{t}_3|=|\vec{t}_3-\vec{t}_4|$) are equal providing constraints for $u(=3/8)$ and $c/a(=\sqrt{8/3})$. The $c/a$ ratio for many WZ phase materials has shown to be close to ideal via X-ray diffraction (XRD), in excellent agreement with predictions from Density Functional Theory (DFT) lattice relaxation calculations. Experimental measurements for the internal structure parameter, $u$, requires measurement of not only the location of the Bragg peaks but also their amplitudes \cite{Kriegner.PRB.13} and has thus proven difficult to extract experimentally. Estimates for $u$ in this work are taken from DFT lattice relaxation calculations (See Table \ref{tab:lat_const_table} for values).

{\setlength{\tabcolsep}{0.55em}
{\renewcommand{\arraystretch}{1.2} 
\begin{table}[]
    \centering
    \begin{tabular}{c| c |c c c}
    &\multicolumn{1}{c|}{ZB} &     \multicolumn{3}{c}{WZ} \\
    Mat.&	 $a_{zb}(\text{\r{A}} )^a$& 	$\frac{a_{wz}}{a_{zb}}-\frac{1}{\sqrt{2}}$&	$\frac{c_{wz}}{a_{wz}}-\sqrt{\frac{8}{3}}$&	 $u-\frac{3}{8}$\\
    \hline					
    AlP& 	5.4635&		    -0.0018$^b$&	    0.0089&	-0.0001\\
    AlAs&	5.6600& 		~0.0019$^b$&	    0.0077&	-0.0003\\
    AlSb&	6.1355& 		~0.0031$^b$&	    0.0059&	-0.0001\\
    GaP& 	5.4509& 	    -0.0021$^b$&	0.0092&	-0.0004\\
    GaAs&	5.6532& 		-0.0021$^b$&	0.0127&	-0.0004\\
    GaSb&	6.0959& 		-0.0028$^b$&	0.0139&	-0.0007\\
    InP& 	5.8687& 		-0.0013$^b$&    0.0078&	-0.0004\\
    InAs&	6.0583& 		-0.0014$^b$&	0.0089&	-0.0005\\
    InSb&	6.4794& 		-0.0014$^b$&	0.0094&	-0.0006\\
    C&   	3.5668& 	    -0.0048$^c$&	0.0273&	-\\
    Si&  	5.4300&     	-0.0073$^d$&	0.0196&	-\\
    Ge& 	5.6563& 	    -0.0021$^e$&	0.0164&	-0.0007\\
    \end{tabular}
    \begin{tabular}{l l l l l l}
    \multicolumn{5}{l}{$^a$Reference \onlinecite{Jancu.prb.1998}~~~~~~~~~~~~~~~~~~~~~~~~~~~~~~~~~~~~~~~~~~~~~~~~~}\\
    \multicolumn{5}{l}{$^b$Reference \onlinecite{Bechstedt.IOP.13}}\\
    \multicolumn{5}{l}{$^c$Reference \onlinecite{DENISOV.DRM.11}}\\
    \multicolumn{5}{l}{$^d$Reference \onlinecite{Kasper.Science.77}}\\
    \multicolumn{5}{l}{$^e$Reference \onlinecite{Rodl.PRM.19}}
    \end{tabular}
    \caption{\small ZB and WZ lattice parameters. For WZ, the differences from ideal ratios are shown. Lattice constants for the WZ III-V's are taken from \textit{ab initio} DFT. The LDL group-IV lattice constants are from XRD. The ideal value of $u$=3/8 was used for C and Si.}
    \label{tab:lat_const_table}
\end{table}
}}
The ZB primitive unit cell is constructed from two interpenetrating fcc lattices, with conventional lattice
\begin{align}
    \vec{a}_1=\frac{a}{2}\{1,1, 0\} &&  \vec{a}_2=\frac{a}{2}\{1,0,1\} && \vec{a}_3=\frac{a}{2}\{0,1,1\}
\end{align}
and basis vectors
\begin{align}
    \vec{t}_1=\{0,0,0\}             & \quad  \vec{t}_2=\frac{a}{4}\{1,1,1\}
\end{align}
where $a$ is the ZB lattice constant (See Table \ref{tab:lat_const_table} for values).

The most obvious difference between the WZ and ZB crystals is the number of atoms in the primitive unit cell. This results in the reciprocal-space volume ($V=(2\pi)^3/|\vec{a}_1\cdot \vec{a}_2\times\vec{a}_3|$) of ideal-WZ ($V_{WZ}$) being half that of ZB ($V_{ZB}$). Consequently, band-folding must occur in order to accommodate the same number of states states in half the (reciprocal-space) volume.

\begin{figure}[ht!]
    \centering
  \begin{subfigure}{1\linewidth}
    \includegraphics[width=1\linewidth]{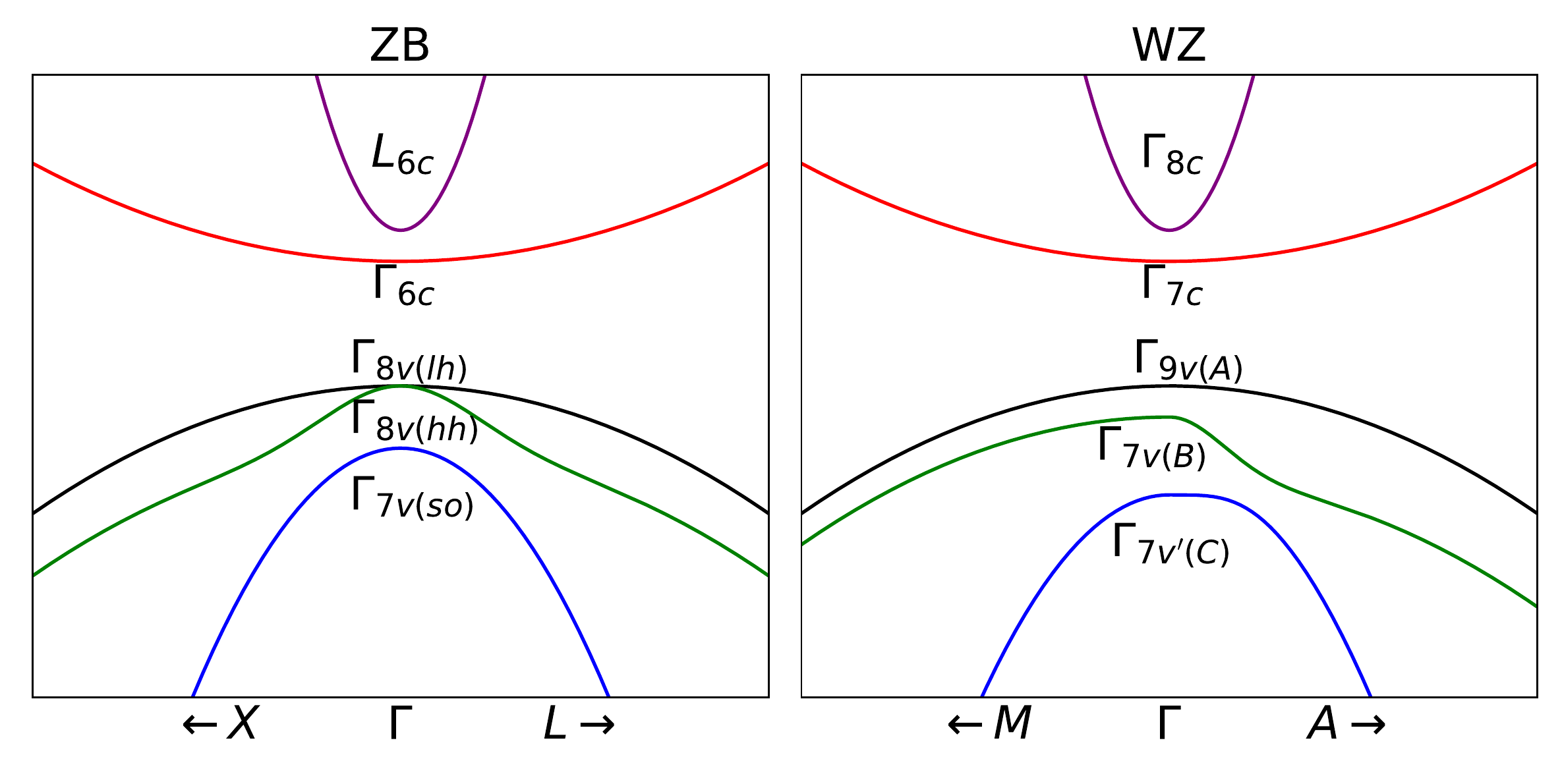}
  \end{subfigure}
\caption[ZB-WZ Band Structure Comparison]{\small Illustrative band diagram for ZB$_4$ and WZ near the $\Gamma$-point. The band folded $L$ point conduction band valley is shown. The ordering of the red and purple bands are material dependent. The ordering shown corresponds to GaAs, InP, InAs and InSb, while AlP, AsAs, AlSb, GaP and GaSb have reversed purple ($L_{6c}/\Gamma_{8c}$) and red bands ($\Gamma_{6c}/\Gamma_{7c}$).} \label{fig:ZB-WZ_bandstructure_comp}
\end{figure}

It is useful to consider the non-primitive ZB unit cell which has been doubled along the $[001]$ direction such that $V_{WZ}=V_{ZB}$. This four atom unit cell, $ZB_4$, has lattice and basis vectors
\begin{align}
    &\vec{a}^\prime_1=\vec{a}_1&& \vec{a}^\prime_2=\vec{a}_{2}&& \vec{a}^\prime_3=2\vec{a}_3
\end{align}
\begin{align}
    &\vec{t}^\prime_1=\vec{t}_1&&   \vec{t}^\prime_2=\vec{t}_2\\
    &\vec{t}^\prime_3=\vec{t}_1+\vec{a}_3&&     \vec{t}^\prime_4=\vec{t}_2+\vec{a}_3.\nonumber
\end{align}
Here, the non-primitive (primed) vectors for $ZB_4$ are expressed in terms of the primitive ZB (un-primed) vectors.
Doubling the ZB unit cell along $[001]$ results in a zone folding of the primitive $L_z$ point  back onto $\Gamma$, allowing half the WZ $\Gamma$-point state to be identified with WZ $\Gamma$-point states, and the other half with  zone-folded states.
This correspondence of states between the WZ and ZB can be used to understand the band ordering of the WZ $\Gamma_{8c}$ and $\Gamma_{7c}$ states (Fig. \ref{fig:ZB-WZ_bandstructure_comp}).

\begin{figure}[ht!]
    \centering
  \begin{subfigure}[b]{0.48\linewidth}
    \includegraphics[width=\linewidth]{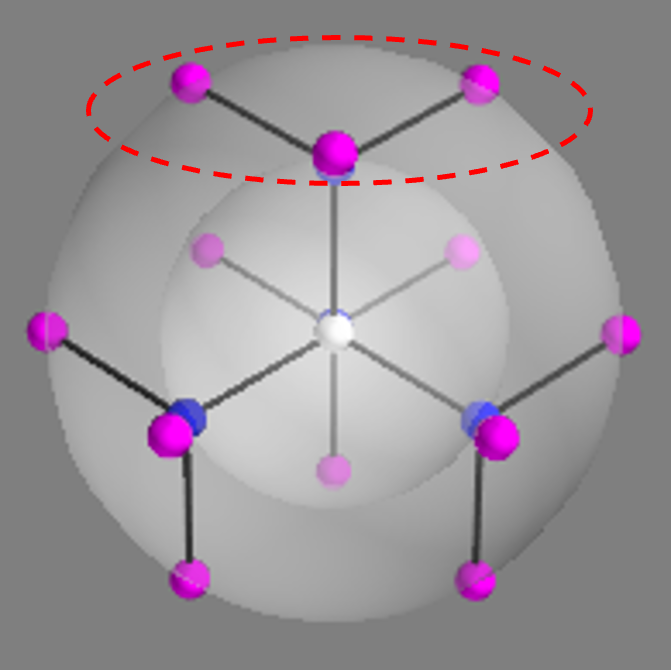}
    \subcaption[ZB with 2$^{nd}$ Nearest Neighbors Side View]{ZB$_4$} \label{fig:geom_3C_111}
  \end{subfigure}%
  \hspace{\fill}   % maximize separation between the subfigures
  \begin{subfigure}[b]{0.48\linewidth}
    \includegraphics[width=\linewidth]{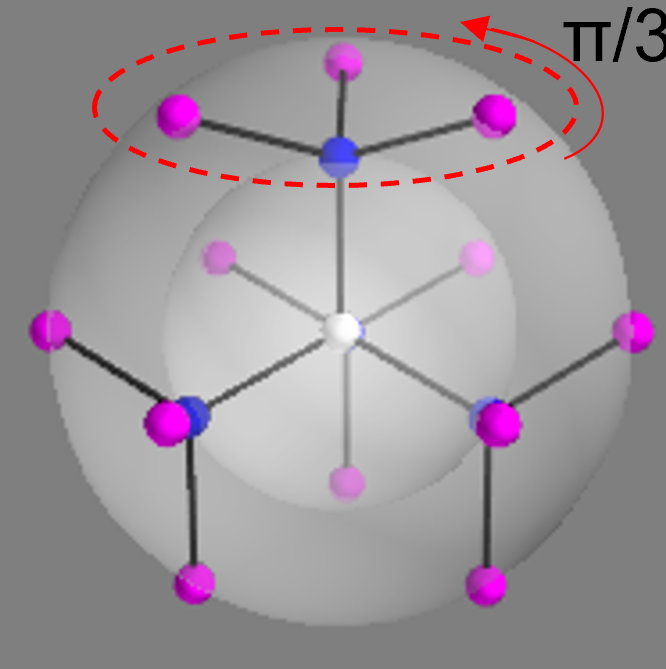}
    \subcaption[WZ with 2$^{nd}$ Nearest Neighbors Side View]{WZ} \label{fig:geom_2H_001}
  \end{subfigure}
  \\
    \centering
  \begin{subfigure}[b]{0.48\linewidth}
    \includegraphics[width=1\linewidth]{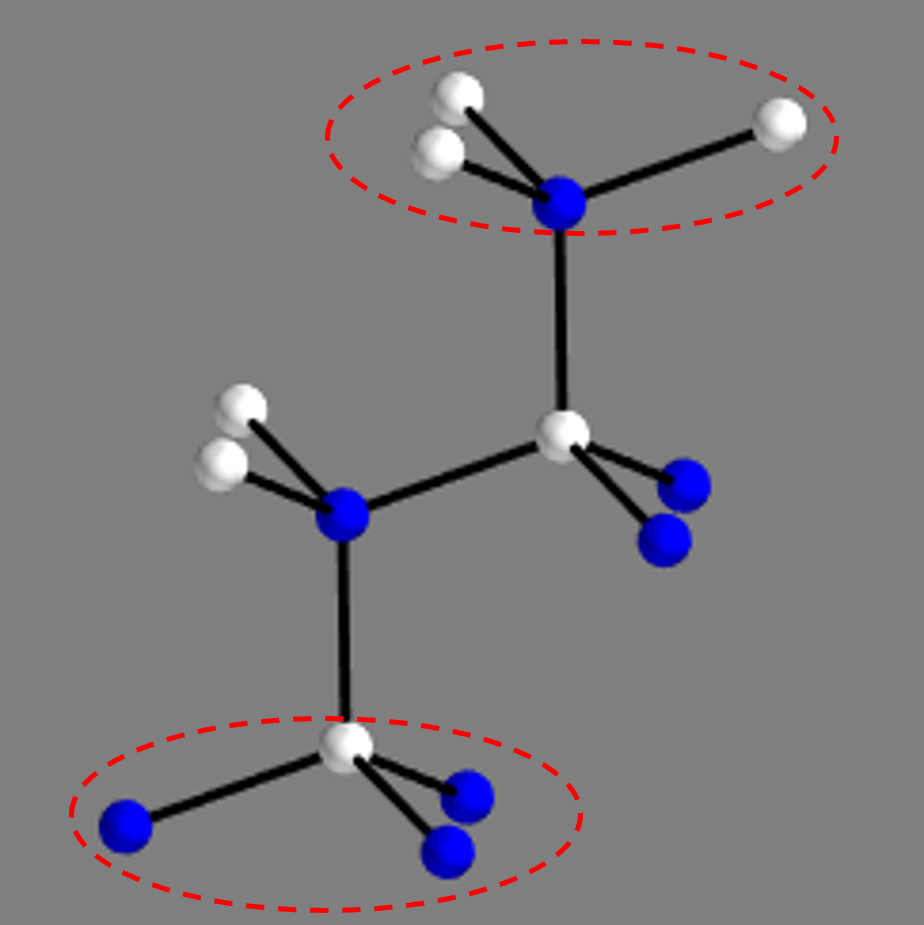}
    \subcaption[Four Atom ZB Unit Cell with Nearest Neighbors]{ZB$_4$} \label{fig:3C_4atom_geom}
  \end{subfigure}
  \hspace*{.5\fill}   
  \begin{subfigure}[b]{0.48\linewidth}
    \includegraphics[width=1\linewidth]{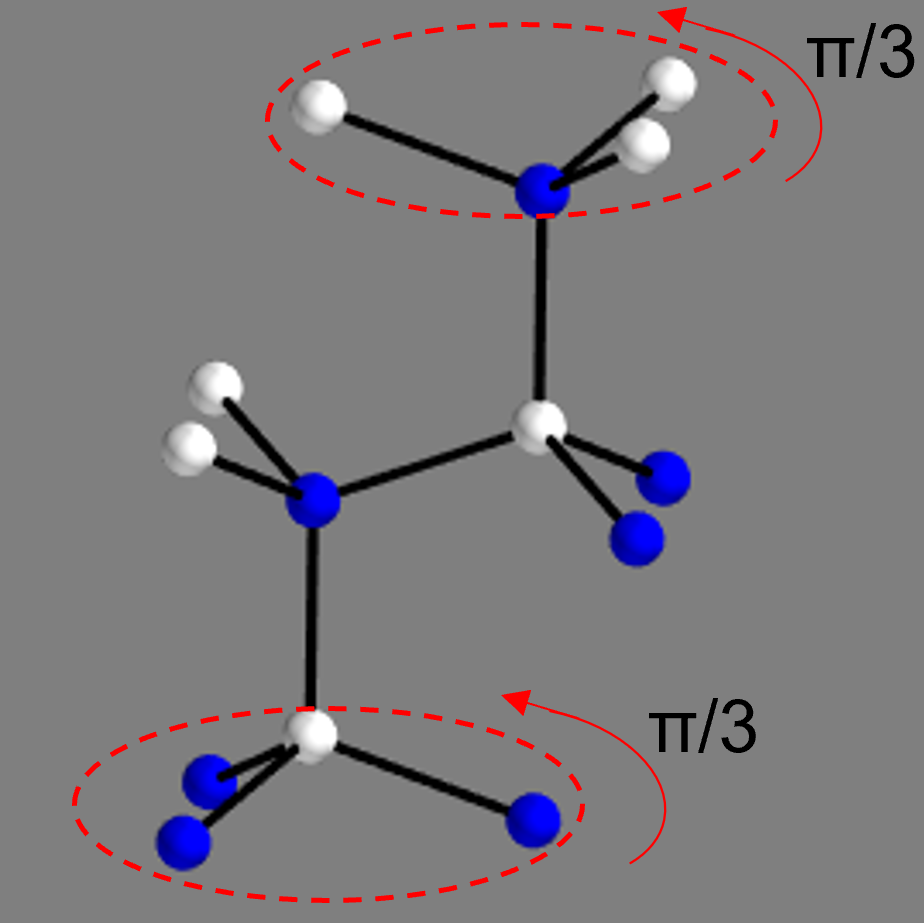}
    \subcaption[WZ Unit Cell With Nearest Neighbors]{WZ} \label{fig:2H_4atom_geom}
  \end{subfigure}
\caption[ZB and WZ with 2$^{nd}$ Nearest Neighbors]{\small Crystal Structure for WZ and ZB. a) ZB viewed along $[111]$, b) WZ viewed along $[0001]$ showing first and second nearest neighbors about a central atom. The white spheres in a) and b) are the central atoms, while the blue and pink spheres are first and second nearest neighbors, respectively. The opaque shells indicate first and second nearest neighbor distances. c) non-primitive four atom ZB unit cell and nearest neighbors, d) primitive WZ unit cell and nearest neighbors. Comparisons of a) c) with b) d) demonstrate the key difference between the WZ and ZB crystals is the three atoms at 2NN distances which are rotated by $\pi/3$.} \label{fig:2NN_geometry}
\end{figure}

At 1NN distances the crystal field environment for the two polytypes are identical, with both materials being tetrahedrally coordinated and having the same bond lengths. At 2NN distances the crystals share a majority of their atomic arrangement with 9 of 12 2NN atoms in the same configuration and the remaining 3 of 12 merely rotated by $\pi/3$ (Fig.\ref{fig:2NN_geometry}). This rotation demotes the crystal symmetry from $T_{d}$ in ZB ($O_h$ in Diamond) to $C_{6v}$ in WZ ($D_{6h}$ in Lonsdaleite). The identical 1NN configuration and extremely similar 2NN configuration serves as the basis to expect a high degree of polytypic transferability of the electronic parameters between the WZ and ZB crystal-phases.

%2H and 3C geometry images 2NN

\subsection{Tight Binding}\label{Hamiltonian}

ETB is a short range atomistic theory that can be used to yield sparse real-space matrices that can be solved efficiently for systems with many atoms \cite{Zielinkski.PRB.12}. Many approximations, some of which are discussed below, are required in order to reduce the number of empirical parameters to a manageable set.

The TB Hamiltonian within the one-electron approximation is expressed as a sum over effective atomic potentials, $V_i(\vec{r})$, centered on atomic sites in the crystal
\begin{align}
    \hat{H}_{1e}(\vec{r})=\hat{T}+\sum_{j,i}\hat{V}_{i}(\vec{r}-\vec{R}_j-\vec{t}_i)+\hat{H}_{so} \label{eq:Ham_1eSO_operator}
\end{align}
where $\hat{T}$ is the kinetic energy operator ($\frac{\hat{p}^2}{2m_e}$), $\vec{R}_j$ is the location of the $j^{th}$ unit cell in the crystal, $\vec{t}_i$ the position of the $i^{th}$ atom in the crystal basis, $V_i$ an effective atomic potential for an atomic species occupying the $i^{th}$ position in the crystal basis ($\vec{t}_i$) and $\hat{H}_{so}$ the spin-orbit interaction. 

To emphasize the bonding nature of the electronic hopping, the ETB basis is typically taken to be constructed from atomic(-like) orbitals centered on the atoms in the unit cell \cite{SlaterKoster.PR.54,Jancu.prb.1998,Tan.PRB.16}
\begin{align}
    &\ket{\psi_{\eta}}=\ket{\phi_{\beta},\vec{R}_{j}+\vec{t}_{i}}\ket{\chi}\label{eq:basis_ket}
\end{align}
where $\eta$ is shorthand for the indices ${\{i\text{(basis)},\ j\text{(unit cell)},\ \beta\text{(orbital)},\ \chi\text{(spinor)}\}}$.

The atomic(-like) orbitals may be taken to be orthogonal between different lattice sites ($\bra{\psi_{\eta}}\ket{\psi_{\eta'}}=\delta_{\eta,\eta'}$) via an orthogonalization procedure such as L\"{o}wdin
or by using Wannier functions. In practice, the orbital wavefunctions are rarely calculated as only knowledge of how they transform is required for most applications. Non-orthogonal ETB models have been shown to increase transferability in some ETB models\cite{Bernstein.PRB.05} at the expense of computational efficiency\cite{Boykin.JAP.19} as the basis requires orthogonalization to resolve ambiguities in various operators (ie position operator for charge localization).

Expanding the one-electron Hamiltonian in Eq. \ref{eq:Ham_1eSO_operator} in an orthogonal (L\"{o}wdin) basis and collecting like terms yields yields

\begin{align}
    &H_{\eta,\eta'}(\vec{r})=&&\left<\psi_{\eta}\left|\hat{T}+\hat{V}_{\eta}(\vec{r})\right|\psi_{\eta'}\right>\delta_{\eta,\eta'} \label{eq:TB_separated_terms}\\
    &&&+\left<\psi_{\eta}\left|\hat{V}_{\eta}(\vec{r})\right|\psi_{\eta'}\right>(1-\delta_{\eta,\eta'}) \nonumber\\
    &&&+\sum_{\eta^{\prime\prime}}
        \left<\psi_{\eta}\left|\hat{V}_{\eta^{\prime\prime}}(\vec{r})\right|\psi_{\eta'}\right>+\left<\psi_{\eta}\left|\hat{H}_{so}\right|\psi_{\eta'}\right>\nonumber
\end{align}
\begin{align}
    &H_{\eta,\eta'}(\vec{r})=&&\mathcal{E}_{\eta}\delta_{\eta,\eta'}+t_{\eta,\eta'}+\sum_{\eta^{\prime\prime}}
            \gamma^{\eta^{\prime\prime}}_{\eta,\eta'}+\Delta^{so}_{\eta,\eta'}\label{eq:H_all_terms_short}
\end{align}
where $\hat{V}_{\eta}(\vec{r})=\hat{V}_{i}(\vec{r}-\vec{R}_j-\vec{t}_i)$ and $\eta$,$\eta^\prime$ and $\eta^{\prime\prime}$ in the 3CI sum correspond to different atomic sites in the crystal. The terms in Eq. \ref{eq:TB_separated_terms} are the onsite-energy $\mathcal{E}_\eta$, two-center integrals (2CI) $t_{\eta,\eta'}$, three-center integrals (3CI) $\gamma^{lm}_{\eta,\eta'}$ and spin-orbit coupling $\Delta^{so}_{\eta,\eta'}$, respectively.

The sum involved in calculating the 3CI's generates a large number of parameters that are typically ignored ($\gamma^{\eta^{\prime\prime}}_{\eta,\eta'}=0$). This truncation is justified post-fact by arguing their influence must be negligible compared to first-nearest neighbor 2CI terms, and reasonable agreement between the two-center model and experimental data remains possible. It has been shown via DFT-TB \cite{QUAMBO.Wang.Springer.2009,Lu.APS.15,Zhang.JPCondMatt.16,Ebihara.PSSC.11} that the inclusion of 3CI and 2CI possibly out to third-nearest neighbor interaction are necessary for polytypic transferability of ETB parameters. The quasi-cubic approximation
\begin{align}
    &\Delta_{xtal},\Delta_{SO}=\frac{1}{2}((\Delta_{AB}+\Delta_{AC})\nonumber\\
    &\pm\sqrt{(\Delta_{AB}+\Delta_{AC})^2-16u\Delta_{AB}\Delta_{AC}})
\end{align}
can be used to estimate the crystal ($\Delta_{xtal}$) and spin-orbit ($\Delta_{SO}$) splittings in WZ materials in terms of the $\Gamma_{9v}$ and $\Gamma_{7v}$ band splittings ($\Delta_{AB}=-(\Gamma_{9v}-\Gamma_{7v}),\ \Delta_{AC}=-(\Gamma_{9v}-\Gamma_{7v'})$).
The deviation from the cubic approximation observed in InP, seen in the differing spin-orbit splittings in the ZB and WZ \cite{Chauvin.APL.18} crystal phases, implies non-negligible contributions from $N^{th}- NN$ and $3CI$ terms.

The Slater-Koster (SK) \cite{SlaterKoster.PR.54} formalism allows for the decomposition of the two-center integrals in terms of products of radial and angular components. The angular dependence is known from the positions of atoms in the structure, while the radial-overlap terms are empirically determined. For small deviations about an ideal geometry, it is common to approximate the unknown scaling of the radial terms with a power scaling law ($\rho\rightarrow \rho (d/d_0)^\beta$) \cite{Jancu.prb.1998} fit to deformation potential data. 

The ZB crystal field splits the 5-fold degenerate d-band into a doublet ($\Gamma_{15}=\{d_{x^2-y^2},d_{3z^2-r^2}\}$) and triplet ($\Gamma_{12}=\{d_{xy},d_{yz},d_{zx}\}$). In the WZ crystal field the p-band splits into a singlet ($\Gamma_1=\{p_z\}$) and doublet ($\Gamma_{6}=\{p_x,p_y\}$), and the d-band into a singlet ($\Gamma_{1}=\{d_{3z^2-r^2}\}$) and two doublets ($\Gamma_{5}=\{d_{xy},d_{x^2-y^2}\}$ and $\Gamma_6=\{d_{yz},d_{zx}\}$). Here, as in Ref. \onlinecite{Jancu.APL.2002}, we ignore the p-band splitting in WZ and represent the d-band onsite energies in terms of the ZB onsite energies as $E_{d_{3z^2-r^2}}=E_{d_{15}}$, $E_{d_{xz}}=E_{d_{yz}}=(E_{d_{12}}+2E_{d_{15}})/3$ and $E_{d_{x^2-y^2}}=E_{d_{xy}}=(2E_{d_{12}}+E_{d_{15}})/3$.

A perturbing field that reduces the crystal point symmetry from the ideal configuration will mix the diagonal onsite energies as
\begin{align}
    E^\prime_i=C_{ij}E_j\label{eq:strain_onsite}
\end{align}
where $E^{\prime}_i$ is the strained onsite energy, $E_i$ is the ideal onsite energy and the $C_{ij}$ is a symmetric matrix whose coefficients are functions of the local strain ($C_{ij}=C_{ij}(\epsilon)$). 

For the ETB model, these coefficients are determined by comparison with deformation potentials. In the presence of shear strain, the d-bands in our model respond via
\begin{align}
    E^\prime_{xy}&=E_{xy}\left(1+2b_d(\epsilon_{zz}-\epsilon_{xx})\right) \label{eq:Shear}\\
    E^\prime_{yz}&=E_{yz}\left(1-b_d(\epsilon_{zz}-\epsilon_{xx})\right)\nonumber\\
    E^\prime_{zx}&=E_{zx}\left(1-b_d(\epsilon_{zz}-\epsilon_{xx})\right)\nonumber\\
    E^\prime_{x^2-y^2}&=E_{x^2-y^2}\qquad \qquad E^\prime_{3z^2-r^2}=E_{3z^2-r^2}\nonumber
\end{align}
where $b_d$ is the shear parameter for the d-bands and $E^\prime$ is the strained onsite energy in terms of the unstrained value for the ideal crystal.

While only the macroscopic strain appears in Eq. \ref{eq:Shear}, an accurate description of both the macroscopic ($\epsilon$) and internal ($\Lambda$) strain tensors are important for describing the valence band states. In AlN, the deviation from ideal $u$ ($u^\prime-u\approx 6.975\ 10^{-  3}$) is responsible for the inversion of the $\Gamma_{9v}$ and $\Gamma_{7v}$ band ordering\cite{Jancu.APL.2002}, although no such band rearrangement occurs among the non-Nitride III-V's. The strain tensors in terms of the real and ideal lattice constants are given by
\begin{align}
    \vec{a}^\prime_i&=(\mathbf{1}+\mathbf{\epsilon})\cdot \vec{a}_i\\
    \vec{t}^\prime_j&=(\mathbf{1}+\mathbf{\Lambda_j}) \cdot (\mathbf{1}+\mathbf{\epsilon}) \cdot \vec{t}_j
\end{align}
The non-zero strain components of $\epsilon$ and $\Lambda_i$ from ideal to the real crystal lattice are $\epsilon_{xx}=\epsilon_{yy}=\frac{a^{\prime}}{a}-1$ and $\epsilon_{zz}=\frac{c^{\prime}}{c}-1$ and $\Lambda_{2,zz}=\frac{u^{\prime}}{u}-1$ and $\Lambda_{4,zz}=\frac{1/2+u^{\prime}}{1/2+u}-1$, respectively.

\subsection{ Nearest Neighbor and 2-Center Approximations}
As noted in Section \ref{Crystals}, WZ and ZB are remarkably similar in their structure out to second nearest neighbors. Importantly, both ideal-WZ and unstrained-ZB are tetrahedrally coordinated, with the deviation of the real WZ crystal from ideal being small ($<1\%$) for most III-V and IV materials. Assuming the first-nearest-neighbor approximation remains valid for the hexagonal phase, WZ and ZB crystals should both equally well described by the same set of ETB parameters.

The 1NN 2CA orthogonal spds$^*$\cite{Jancu.prb.1998} model is the simplest model that can describe well simultaneously the valence band and lowest conduction band states, including the location of the conduction band minima. As the first nearest neighbor bond lengths in ideal-WZ and unstrained-ZB are by definition the same, the radial overlap parameters transfer from ZB to WZ intact.

\begin{figure}[ht!] 
	\centering 
	\begin{subfigure}[b]{1\linewidth}
	\centering 
	\includegraphics[width=1\linewidth]{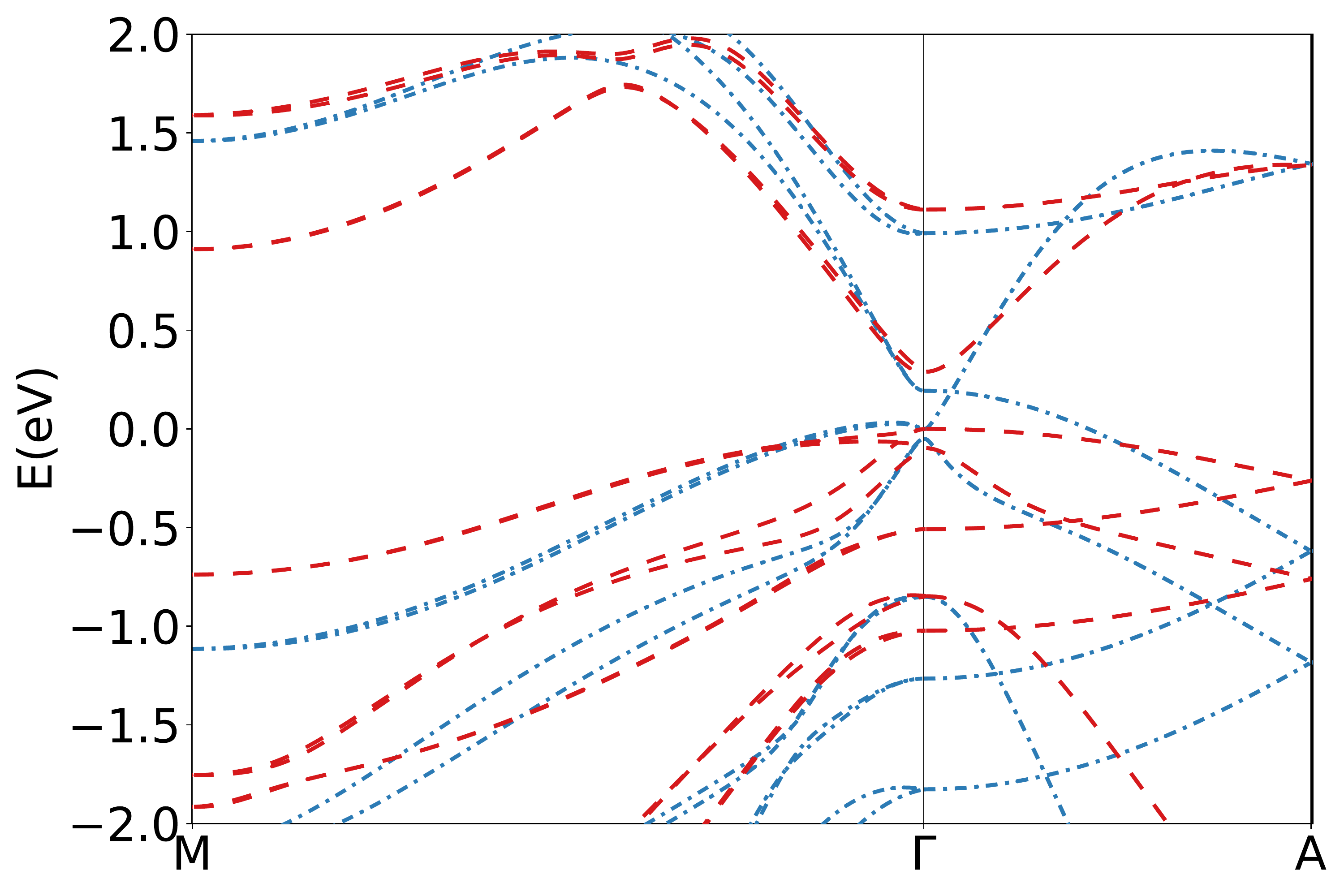}
	\subcaption[Failure of Direct Transferability of TB Parameters: InSb]{\label{fig:Naive_Transferability_bands_InSb}WZ InSb}
	\end{subfigure}
	\caption[Failure of Direct Transferability of TB Parameters]{\small \label{fig:Naive_Transferability_bands} Electronic band structures for InSb calculated using Slater-Koster parameters fit to ZB taken from Ref. \onlinecite{Jancu.prb.1998} are shown in blue (dashed-dotted). Red (dashed) curves are EPP band structures taken from Ref. \onlinecite{De.IOP.2014}. Assuming polytypic transferability between WZ and ZB within the ETB model results in WZ-InSb to become metallic. Band structures calculated with EPP are sufficiently transferable such that only modest changes to the WZ-InSb band gap are observed, as expected.}
\end{figure}

Some results for the `naive transferability' attempt are shown in Figs. \ref{fig:Naive_Transferability_bands}. The largest observed error and general trend in these plots are the systematic underestimation of the WZ band structure as a result of the top of the valence band ($\Gamma_{9v}$) migrating upward in energy. For the extreme case of InSb, this underestimation leads to the `na\"ive'-model predicting a transition to a metal in the WZ phase.

The failure of the model parameters to be directly transferable between the ZB and WZ phases is not all-together unexpected, and is likely due to the effects of second nearest and/or three-center interactions inadvertently being compensated for by the remaining 1NN hopping elements during optimization. This compensation is necessary to describe the ZB band structure in the truncated basis, but consequently causes the model to fail to transfer to the WZ configuration. 

While strict transferability of the ETB parameters generates obviously incorrect results, only moderate corrections to the parameters (reflecting the subtle changes in long range interactions) are required to correct many of these discrepancies. We propose a new set of ETB parameters empirically fit to available WZ experimental and theoretical empirical pseudo-potential (EPP) data \cite{De.PRB.2010,De.IOP.2014}, subject to the constraint that they remain close to the ZB reference model in Ref. \onlinecite{Jancu.prb.1998}. Polytype transferability is accomplished by starting with the ZB parameterization and allowing the electronic parameters to deviate slightly to yield better agreement with experimental and theoretical data.

\section{Methods}\label{sec:methods}
The parameters were fit by minimizing a weighted least squares objective function constructed from the absolute fractional errors with the fitting targets. The objective function was
\begin{center}
    \begin{align}
        \chi(\vec{p})=\sum^{N}_{i=1} W_i\left|\frac{t_i-f_i(\vec{p})}{t_i}\right|^2\label{eq:weighted_least_squares}
    \end{align}
\end{center}
where $\vec{p}$ is the vector of parameters, $t_i$ is a target value, $f_i(\vec{p})$ is a function of the computed band-structure corresponding to $t_i$ and $W_i$ are weights. $W_i$ allows for control of the importance a particular target (such as the band gap energy or a valley mass) has during optimization. 

Many important electronic properties of semiconductors are governed by the energy landscape for electrons and holes at or near the band-gap. Consequently, properties such as valley states, band-curvatures and energy splittings receive a great deal of attention in the description of electronic states of semiconducting materials. In general, states close to the gap receive greater emphasis (weighted higher) compared to states more remote in energy since transport and optical properties are determined largely by the energy landscape at the bandgap. Particular attention is given to direct and indirect bandgaps, energy splittings (ie $\Delta_{ab},\Delta_{ac}$), and effective masses.

To enforce the `closeness' constraint of the ETB-parameters to the ZB reference model for maximal transferability, a quadratic penalty function was used 
\begin{center}
    \begin{align}
    \mathcal{P}(\vec{p})=W_0\sum_i\left|\frac{(p_i-p_{i,0})}{p_{i,0}}-\delta\right|^2\label{eq:quad_penalty_function}
    \end{align}
\end{center}
where $\delta$ and $W_0$ control the effective width of the confining well centered on the vector $\vec{p}_0$. Confinement in this way proved effective in preventing `parameter wonder' during fitting. The inherent non-uniqueness of ETB models is a consequence of the large number of degrees of freedom relative to the number of constraints. In particular, parameters involving the s$^*$ band are nearly unconstrained. They are kept as they improve the transferability of the s bands. A confining weight and width of $W=.1$ and $\delta=.05$ were used during optimization.

Since we expect small differences between the WZ and ZB parameters, $\vec{p}_0$ for each WZ material was taken to be the corresponding ZB parameterization from Ref. \onlinecite{Jancu.prb.1998}. We initialize the d and s$^*$ onsite energies with parameters from Ref. \onlinecite{Jancu.prb.1998}, in which a single value was used
\begin{align}
    E_{d_{12},a}=E_{d_{15},a}&=E_{d_{12},c}=E_{d_{15},c}\label{eq:jancu_d_s-star_simplification}\\
    E_{s^{*},a}&=E_{s^{*},c}\nonumber.
\end{align}
The parameters in Eq.\ref{eq:jancu_d_s-star_simplification} were then allowed to split during optimization, as is done for the WZ III-Nitrides in Ref. \onlinecite{Jancu.APL.2002}. The ETB-parameters were then optimized to be in general agreement with available experimental and theoretical data \cite{De.PRB.2010,De.IOP.2014} while maintaining maximal closeness with the ZB polytype phase enforced via Eq.\ref{eq:quad_penalty_function}. The WZ model values and ZB deviations can be found in Tables \ref{tab:2H-IV-params}, \ref{tab:III-V-series-2H_params} and the corresponding comparisons with target values in Tables \ref{tab:Al_target_values}, \ref{tab:Ga_target_values}, \ref{tab:In_target_values}.

Limiting the fitting to only small corrections, it was found to be only possible to reproduce the direct gap and lowest remote valley. The WZ K-point proved difficult to reproduce while simultaneously maintaining the desired level of transferability with the ZB parameters. Accurate fitting of the WZ-K conduction band valley minima strongly perturbs the ZB-L point conduction band minima as well as in general causing significant deviation of the parameters from the ZB values. As the WZ-K valley is generally remote in energy and in k-space, a general preference of WZ-ZB transferability and direct gap over enforcement of remote targets was adopted.

\section{Results and Discussion}
\label{sec:ResultsAndDiscussion}
\begin{figure}[ht!] 
	\centering
%IV-series bandstructure
	\begin{subfigure}[b]{.32\linewidth}
	\includegraphics[width=1\linewidth]{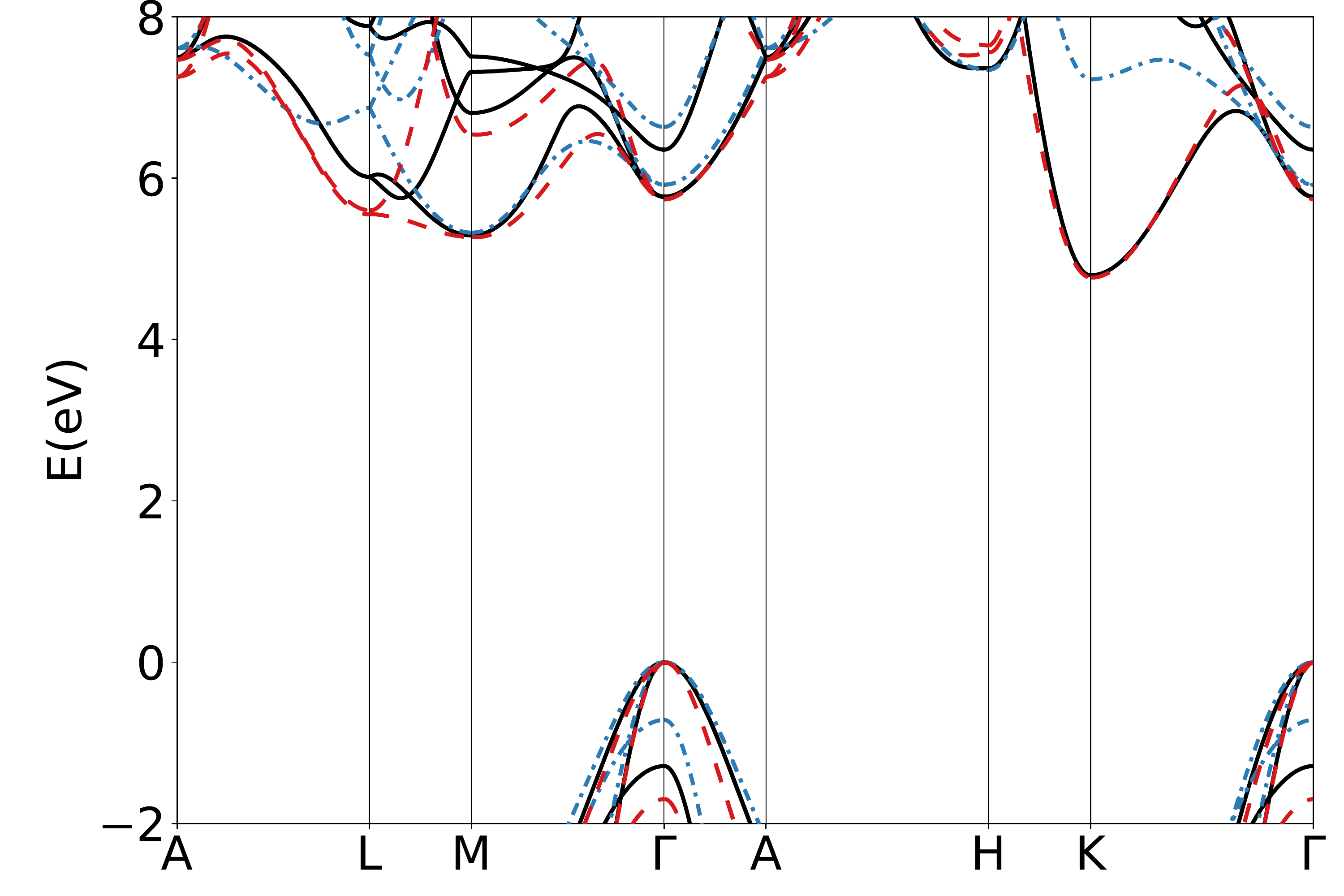}
	\subcaption[]{C}
	\end{subfigure}
    \hspace{\fill}
	\begin{subfigure}[b]{.32\linewidth}
	\includegraphics[width=1\linewidth]{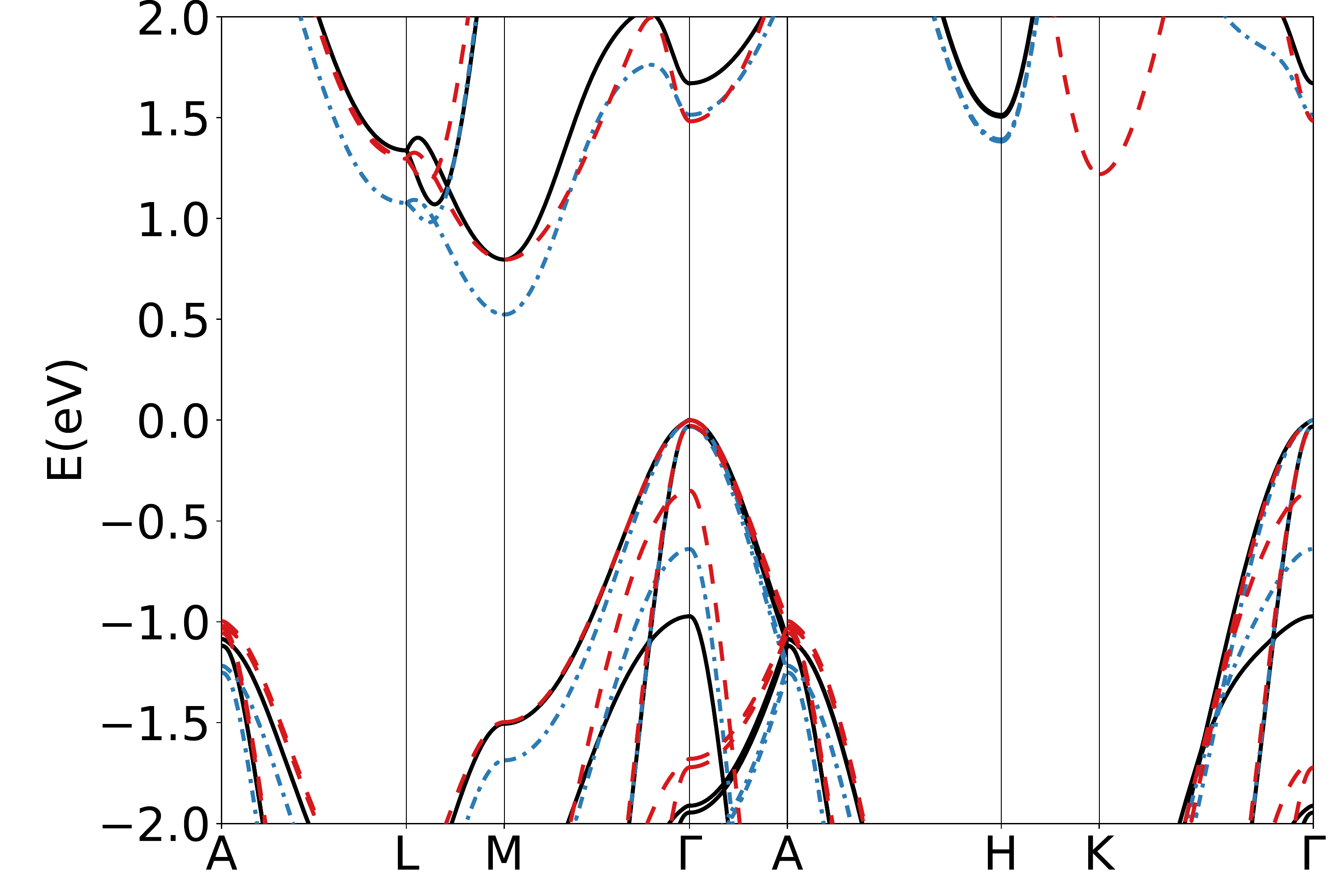}
		\subcaption[]{Si}
	\end{subfigure}
    \hspace{\fill}
	\begin{subfigure}[b]{.33\linewidth}
	\includegraphics[width=1\linewidth]{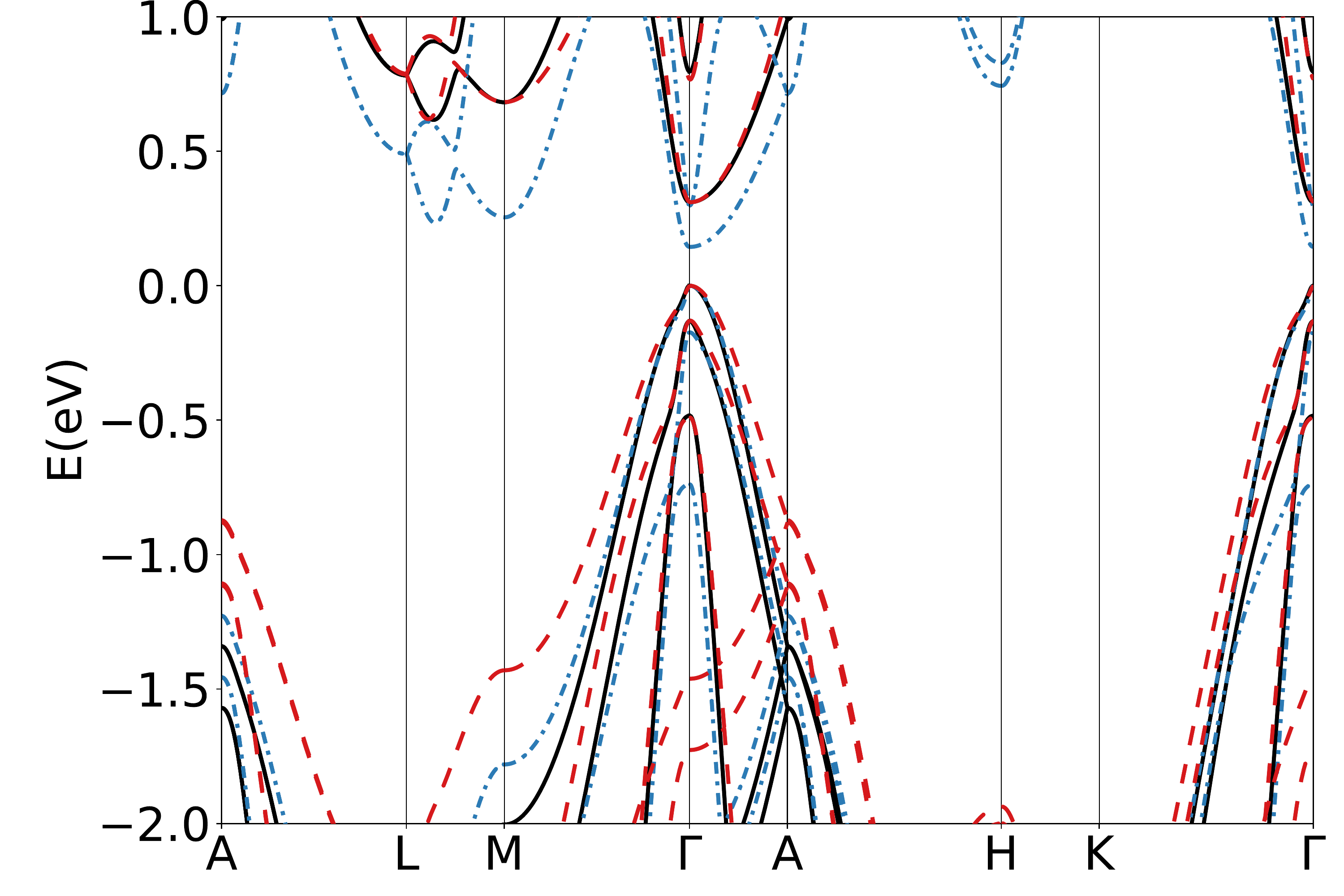}
		\subcaption[]{Ge}
	\end{subfigure}
\\
%Al-series bandstructure
	\begin{subfigure}[b]{.32\linewidth}
	\includegraphics[width=1\linewidth]{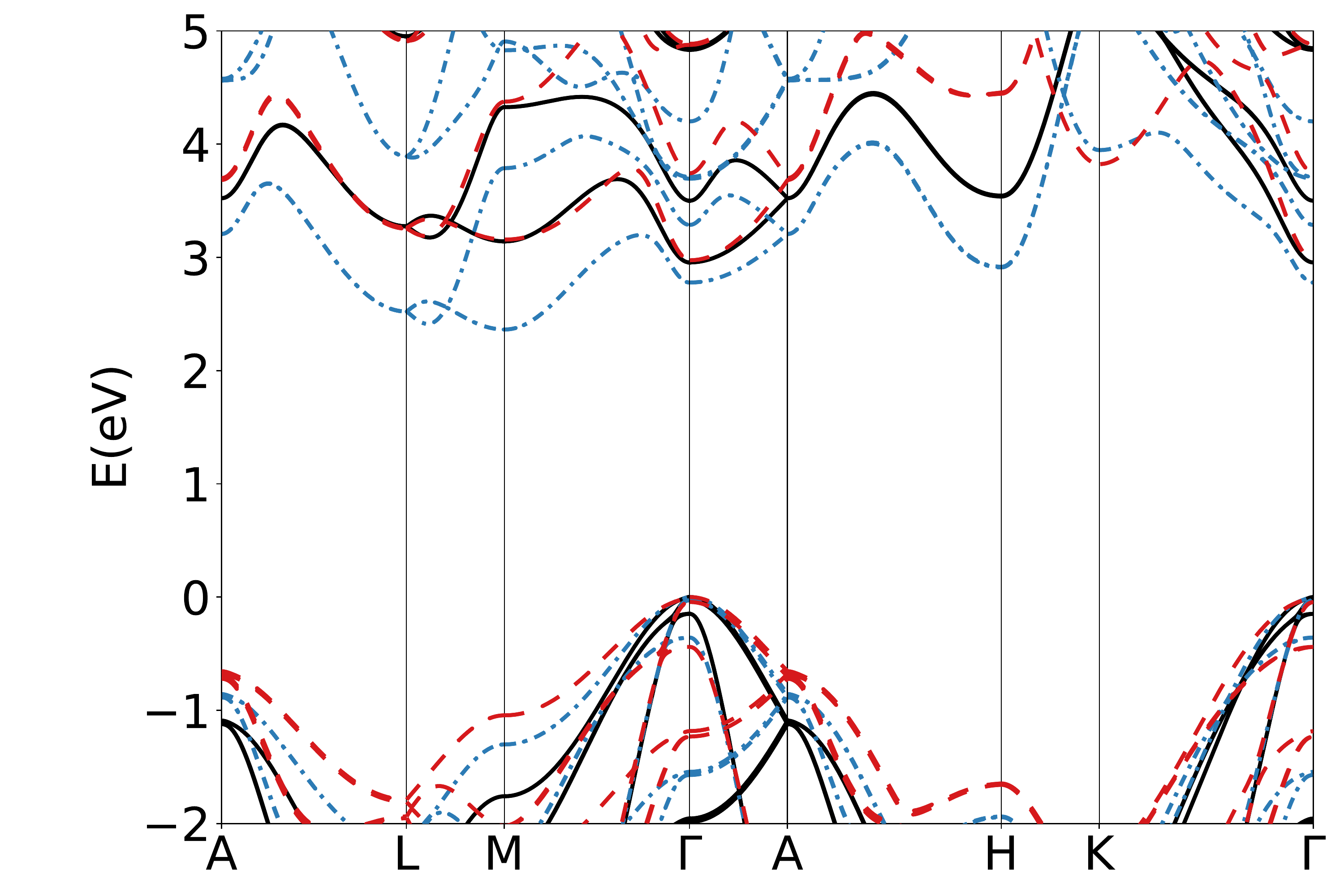}
	\subcaption[]{AlP}
	\end{subfigure}
    \hspace{\fill}
	\begin{subfigure}[b]{.32\linewidth}
	\includegraphics[width=1\linewidth]{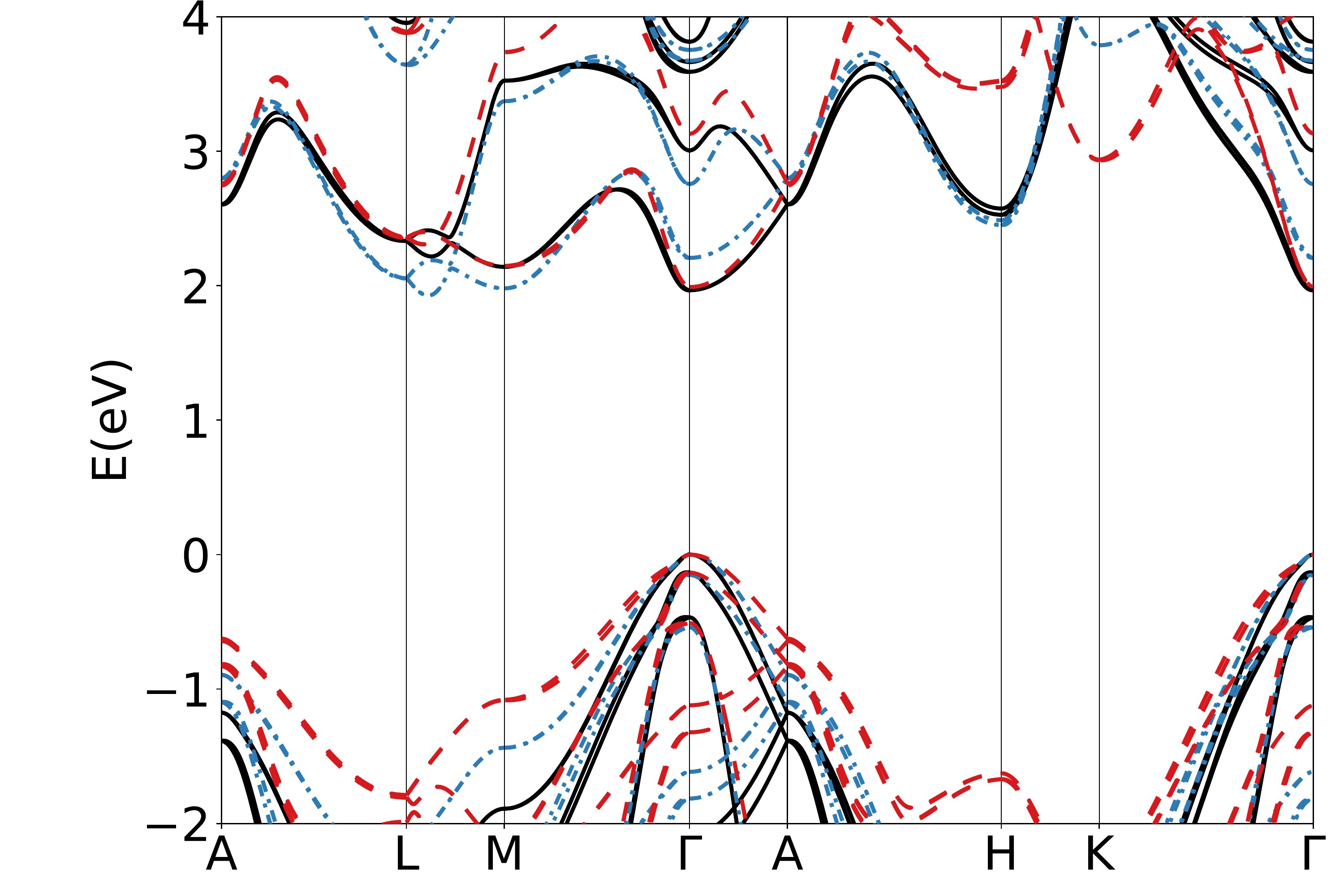}
		\subcaption[]{AlAs}
	\end{subfigure}
    \hspace{\fill}
	\begin{subfigure}[b]{.32\linewidth}
	\includegraphics[width=1\linewidth]{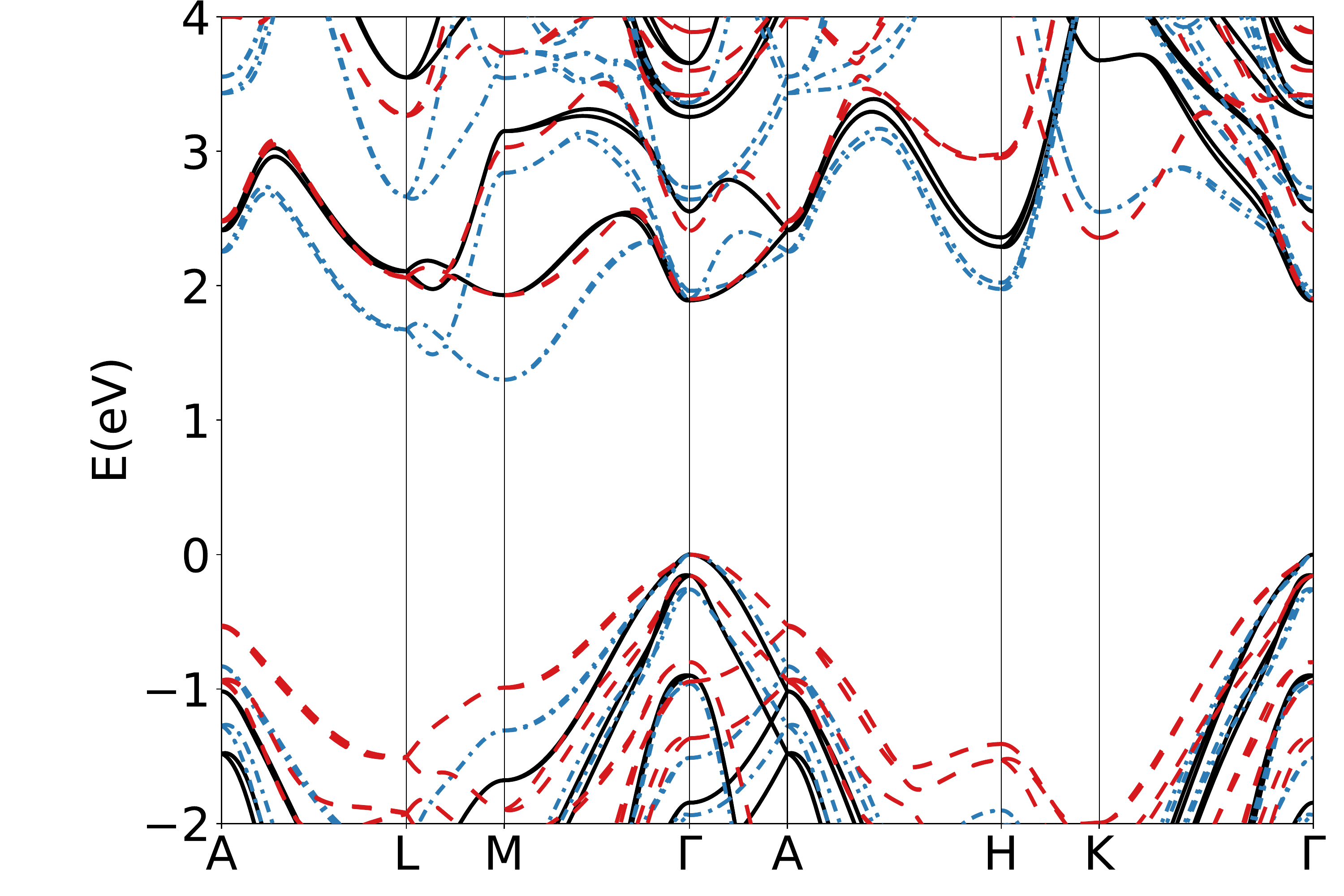}
		\subcaption[]{AlSb}
	\end{subfigure}
\\
	\begin{subfigure}[b]{.32\linewidth}
	\includegraphics[width=1\linewidth]{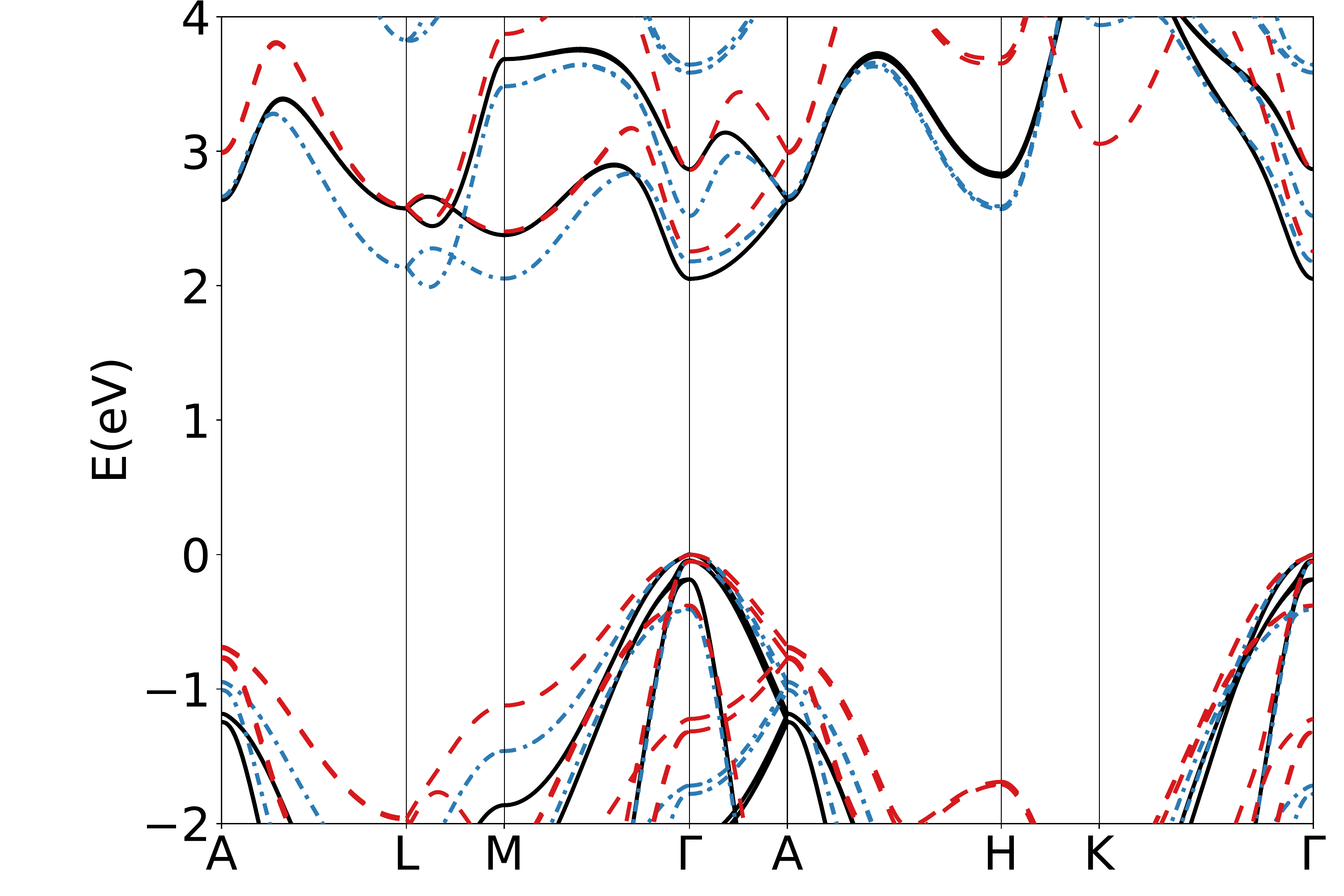}
	\subcaption[]{GaP}
	\end{subfigure}
    \hspace{\fill}
	\begin{subfigure}[b]{.32\linewidth}
	\includegraphics[width=1\linewidth]{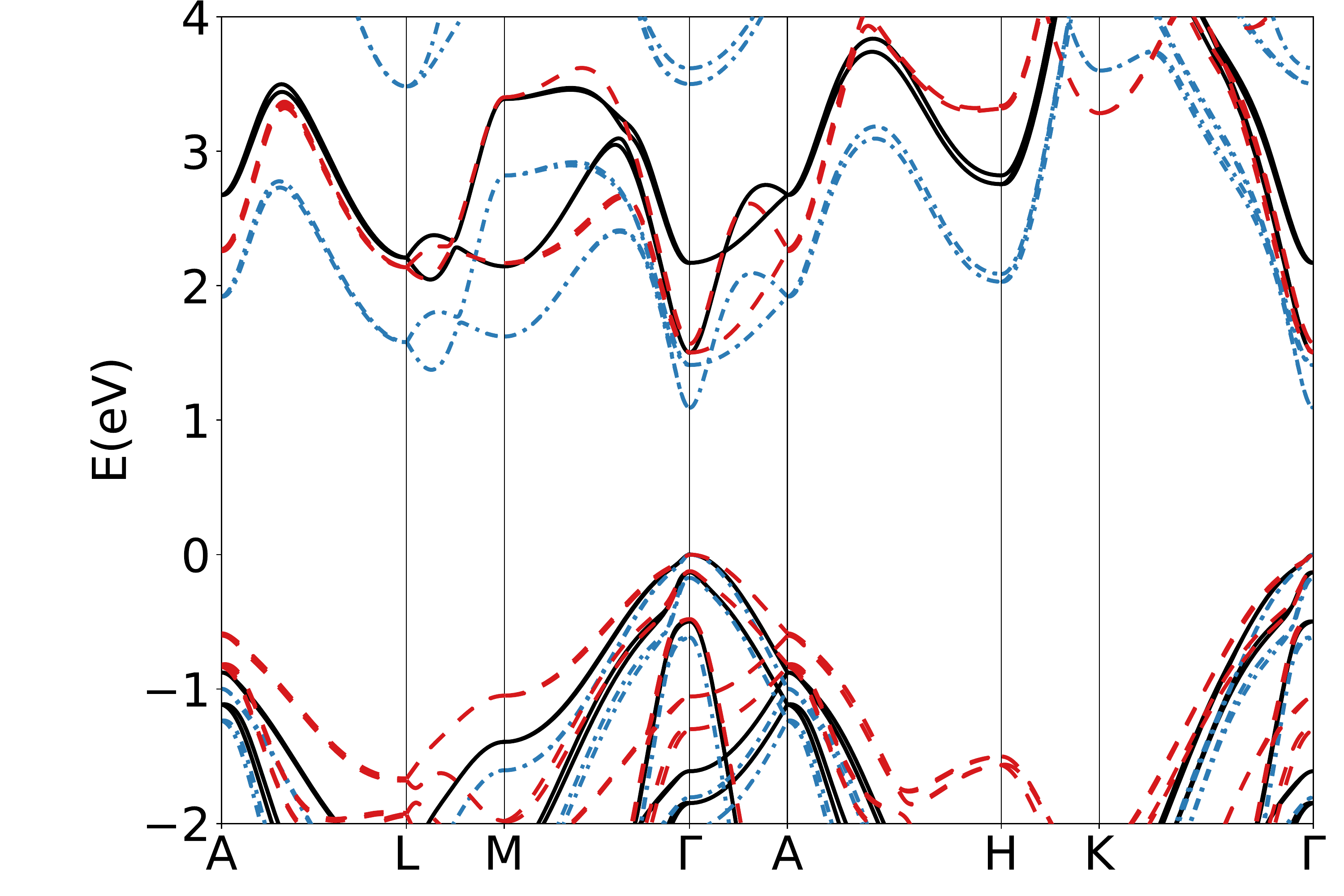}
		\subcaption[]{GaAs}
	\end{subfigure}
    \hspace{\fill}
	\begin{subfigure}[b]{.32\linewidth}
	\includegraphics[width=1\linewidth]{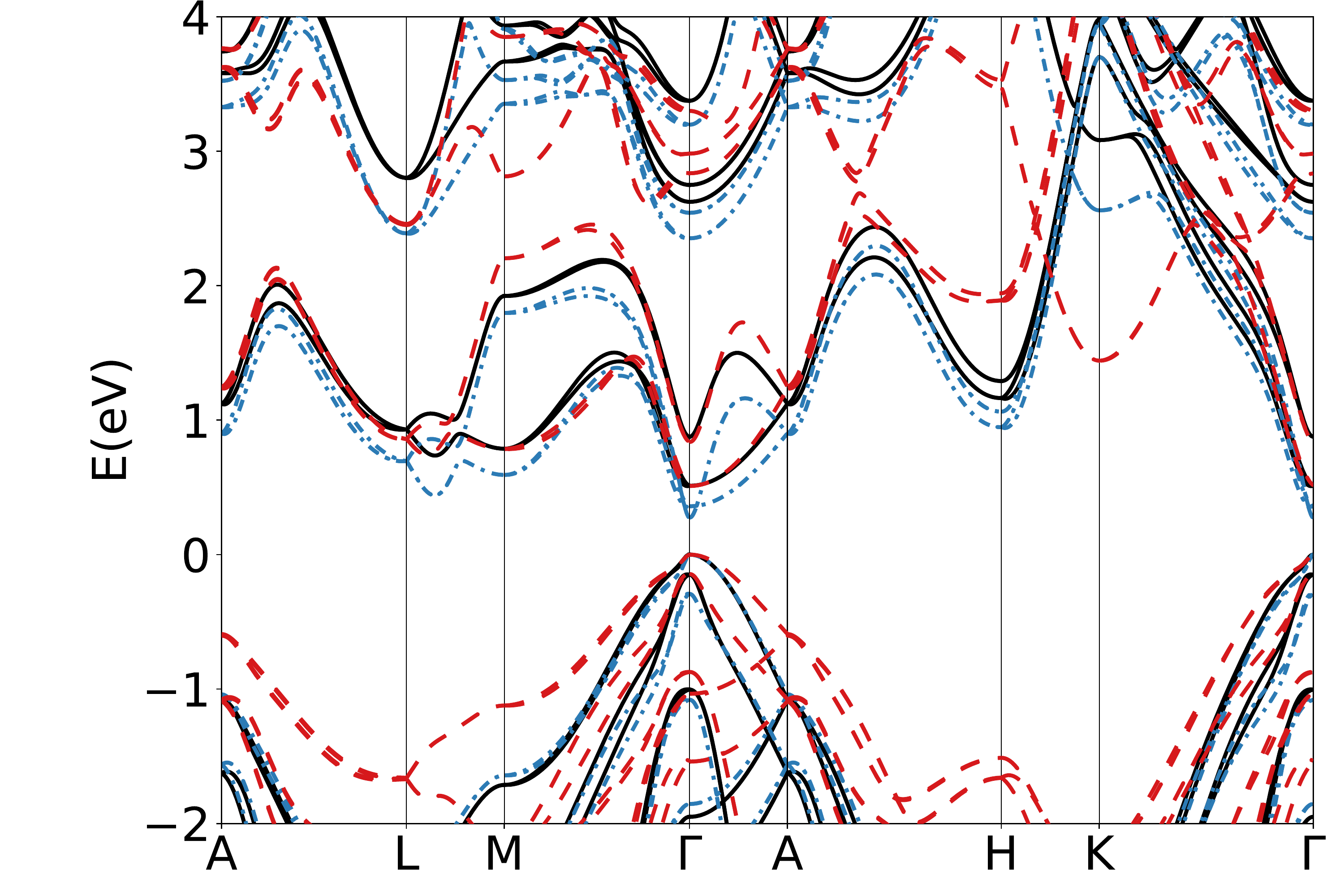}
		\subcaption[]{GaSb}
	\end{subfigure}
\\
	\begin{subfigure}[b]{.32\linewidth}
	\includegraphics[width=1\linewidth]{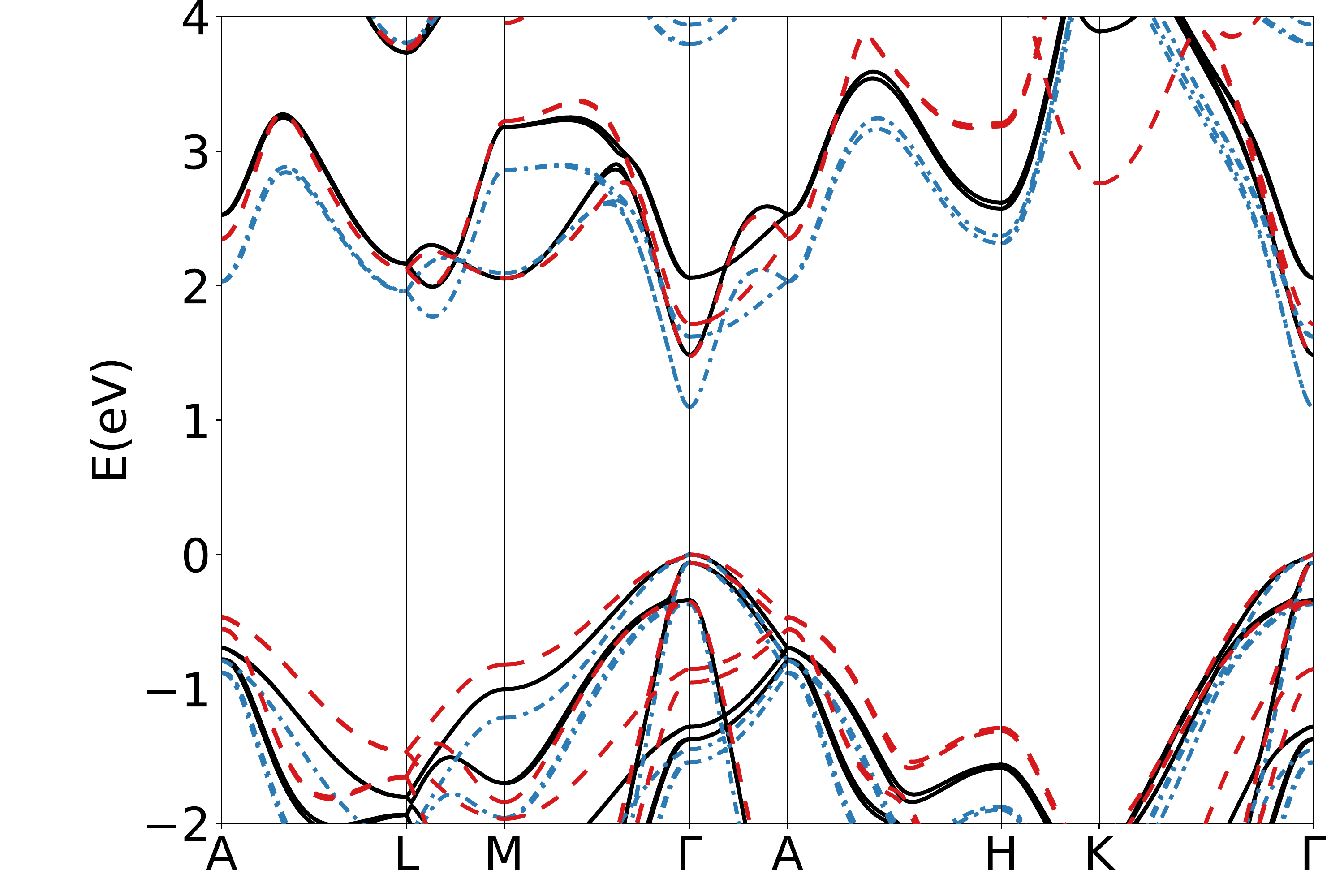}
	\subcaption[]{InP}
	\end{subfigure}
    \hspace{\fill}
	\begin{subfigure}[b]{.32\linewidth}
	\includegraphics[width=1\linewidth]{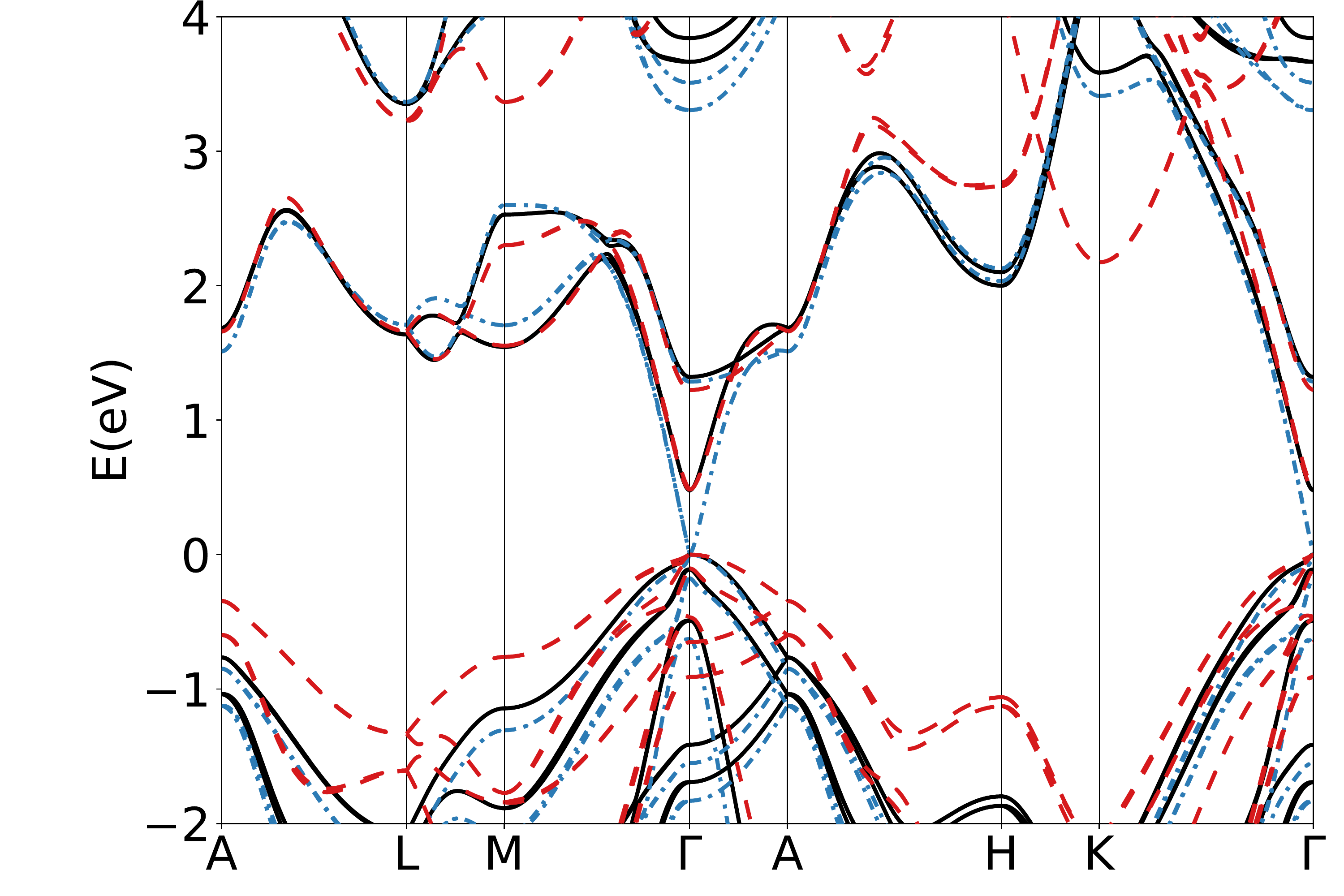}
		\subcaption[]{InAs}
	\end{subfigure}
    \hspace{\fill}
	\begin{subfigure}[b]{.32\linewidth}
	\includegraphics[width=1\linewidth]{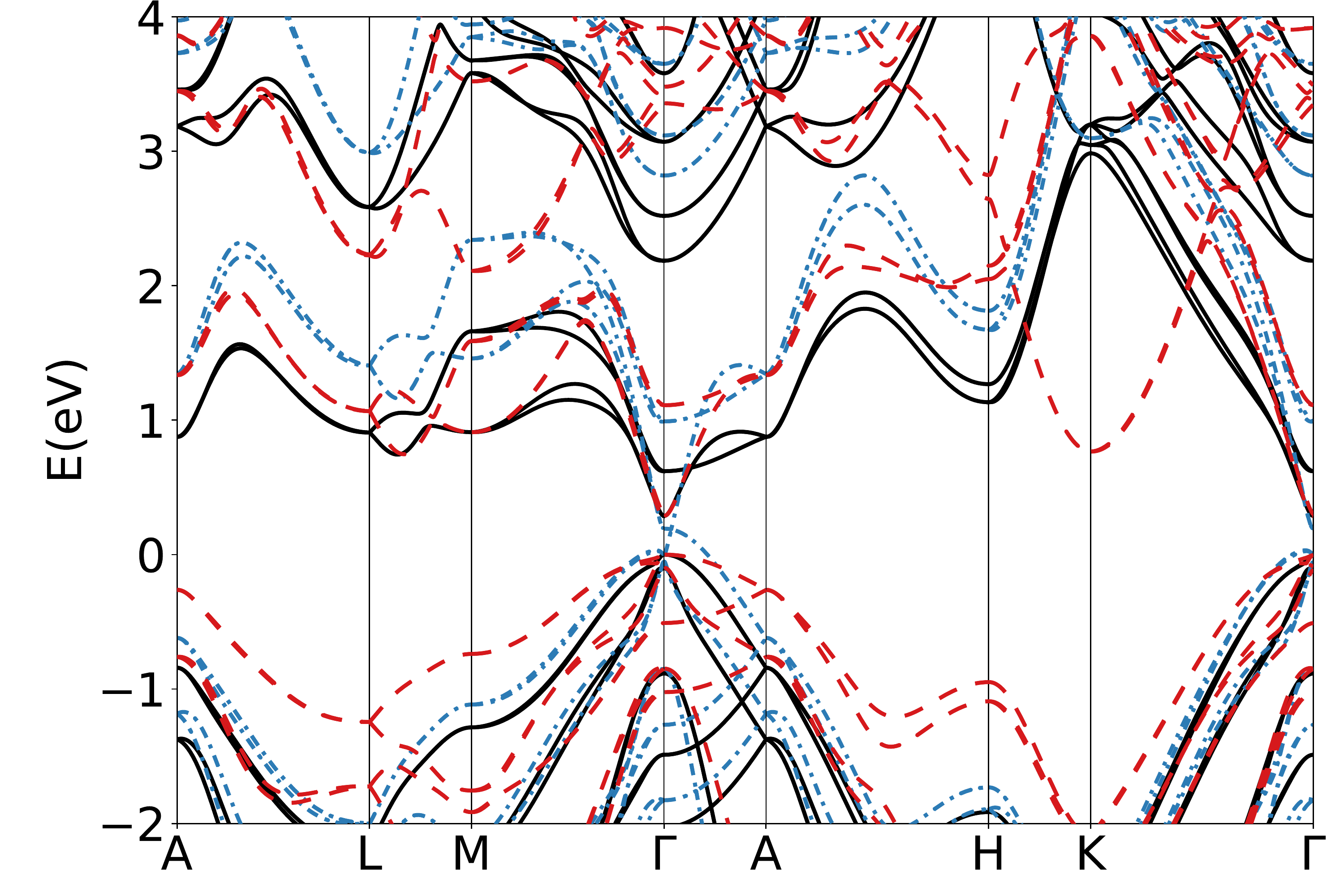}
		\subcaption[]{InSb}
	\end{subfigure}
	\caption[WZ Band Structures]{\small Electronic band structures for the ideal WZ crystal phase (Table \ref{tab:lat_const_table}). The current WZ optimized model is shown as a solid black line, and the ZB optimized ETB parameters from Ref. \onlinecite{Jancu.prb.1998} are shown as a blue dashed-dotted lines. EPP band structures from Refs. \onlinecite{De.IOP.2014,De.PRB.2010} are shown as red dashed lines.}\label{fig:2H-plots}
\end{figure}

All models are now in close agreement with the expected bandgap energies and locations of the conduction band minima. Parameters and comparisons for the WZ group III-V and IV materials can be found in Tables \ref{tab:2H-IV-params}, \ref{tab:III-V-series-2H_params} and Tables \ref{tab:IV_target_values}, \ref{tab:III-V-comparison-tables}, respectively, and corresponding band structures in Figs. \ref{fig:2H-plots}. 

Notably, InSb is no longer metallic and InAs , as was the case in with the na\"ive transferability. All materials now have the expected band symmetry at the gap and location of the conduction band minima. Most materials are direct band-gap materials in the WZ phase with the exception of C and Si which have indirect minima at K and M, respectively. 
~
{\setlength{\tabcolsep}{0.2em}
\renewcommand{\arraystretch}{1.1} 
\begin{table}[ht!]
    \centering
    \begin{tabular}{c|cc|cc|cc}
  ~	&	\multicolumn{2}{c|}{C}&\multicolumn{2}{c|}{Si}&\multicolumn{2}{c}{Ge}\\ 
    \hline
    \hline
    $\lambda$	    &	~0.0030	&( 0.00)&	~0.0193	&( 0.01)&	~0.1301	&( 0.02)\\ 
    \hline
    $E_{s}$	        &	-0.6252	&( 0.01)&	-2.5797	&(-0.02)&	-3.8222	&(-0.01)\\ 
    $E_{p}$	        &	7.0591	&( 0.06)&	4.2663	&(-0.05)&	4.0307	&( 0.03)\\ 
    $E_{d_{12}}$	&	28.8484	&(-0.02)&	14.3263	&(-0.05)&	12.0352	&( 0.04)\\ 
    $E_{d_{15}}$	&	23.4616	&( 0.17)&	11.995	&( 0.12)&	13.7182	&(-0.09)\\ 
    $E_{s^*}$	    &	38.8847	&(-0.01)&	19.2604	&(-0.01)&	18.8331	&(-0.01)\\ 
    \hline
    $s s \sigma$	&	-4.3234	&( 0.01)&	-1.8439	&( 0.05)&	-1.5651	&(-0.04)\\ 
    $s p \sigma$	&	~6.5614	&(-0.19)&	~2.8499	&(-0.02)&	~2.7295	&( 0.02)\\ 
    $s d \sigma$	&	-2.4101	&( 0.13)&	-2.6879	&( 0.04)&	-2.881	&(-0.03)\\ 
    $s s^* \sigma$	&	-2.4022	&(-0.01)&	-1.7396	&(-0.02)&	-1.8272	&( 0.05)\\ 
    $p p \sigma$	&	~8.4125	&(-0.11)&	~4.1471	&(-0.01)&	~4.5227	&(-0.06)\\ 
    $p p \pi$	    &	-2.9551	&(-0.12)&	-1.5099	&( 0.05)&	-1.6515	&( 0.00)\\ 
    $p d \sigma$	&	-2.3075	&(-0.07)&	-2.1762	&(-0.03)&	-2.2941	&(-0.04)\\ 
    $p d \pi$	    &	~4.2872	&(-0.09)&	~1.9477	&( 0.03)&	~1.6439	&( 0.13)\\ 
    $s^* p \sigma$	&	~5.0475	&( 0.02)&	~2.7509	&( 0.03)&	~2.8755	&(-0.02)\\ 
    $d d \sigma$	&	-3.2526	&( 0.22)&	-1.2144	&( 0.01)&	-1.2208	&( 0.00)\\ 
    $d d \pi$	    &	~5.9774	&(-0.20)&	~2.6039	&(-0.04)&	~2.4763	&( 0.01)\\ 
    $d d \delta$	&	-3.9583	&( 0.01)&	-2.4427	&( 0.01)&	-1.9936	&( 0.07)\\ 
    $s^* d \sigma$	&	-2.3285	&(-0.01)&	-0.7057	&(-0.01)&	-0.6200	&( 0.00)\\ 
    $s^* s^*\sigma$	&	-2.6714	&( 0.00)&	-3.2861	&( 0.01)&	-3.6254	&(-0.01)\\ 
    \hline
    $E_{n.i.}$&-0.0757&-&-0.0790&-& -0.0089&-\\
    \end{tabular}
    \caption{\small LDL group-IV SK-parameters in eV. Values in parenthesis are the fractional difference between our LDL parameters and the DMD parameters from Ref. \onlinecite{Jancu.prb.1998}. The onsite energies have been shifted to make the top of the valence band have zero energy ($E(\Gamma_{9v})=0$) for the ideal WZ crystal (Table \ref{tab:lat_const_table}). $\Delta E_{n.i.}=E'(\Gamma_{9v})-E(\Gamma_{9v})$ is the shift in the valence band maximum when going from ideal to non-ideal WZ. To obtain the parameters for the non-ideal structure, subtract $\Delta E_{n.i}$ from the ideal onsite energies.}
    \label{tab:2H-IV-params}
\end{table}
}

While there is insufficient deformation data available to re-fit the radial scaling coefficients for WZ, the similarity between the ZB and WZ ETB-parameters (Table \ref{tab:2H-IV-params}, \ref{tab:III-V-series-2H_params}) as well as the character of states near the gap improves the likely hood that the the radial scaling coefficients will be transferable between the ZB and WZ phase. In Table \ref{table:def_pot_comp} we compare the pressure coefficients for the ZB phase and the WZ phase for similar states near the band gap. The ZB pressure coefficients were calculated using the ZB optimized Slater-Koster and radial-scaling coefficients from Ref. \onlinecite{Jancu.prb.1998}. The WZ pressure coefficients were calculated using the WZ optimized parameters from the present work and the radial-scaling coefficients from Ref. \onlinecite{Jancu.prb.1998}. Many of the pressure coefficients are similar for the WZ and ZB phase, in qualitative agreement with results from Ref. \onlinecite{Wei.PRB.99} where it was found that only small variations between the ZB and WZ deformation potentials were present for the III-V nitrides.

\section{Conclusion}
We have demonstrated some of the problems that arise when using ETB parameterizations determined for bulk ZB in a WZ configuration, despite the high degree of geometric similarity between these polytypes. The failure of the ETB parameters to be transferable between polytypes is a direct consequence of approximations made in the 1NN 2CA model. These approximations result in a failure to accommodate subtle changes in the bonding environment which are observed starting in the three-center integrals and the second nearest neighbor two center integrals. We have produced a set of ETB parameters for WZ non-nitride III-V's and group IV materials that are in significantly improved agreement with available experimental data as well as calculations performed using empirical pseudopotentials. Special attention has been given to the direct gap and next lowest conduction band valley states, as well as transferability with the ZB parameters. The differences between the WZ and ZB ETB parameters are the necessary corrections to accommodate for changes in the crystal field. 
The addition of the ETB parameters presented in this work, optimized for the hexagonal 2H (WZ) phase and transferability with existing ETB models for cubic 3C (ZB) III-V (non-nitride) and group-IV materials, enables the investigation heterostructures containing hexagonal and 3C polytypes.

\onecolumngrid
~
{\renewcommand{\arraystretch}{1.1} 
\begin{sidewaystable}[ht!]
    \small
    \centering
    \begin{tabular}{c||  cc| cc| cc|   cc| cc| cc|   cc| cc| cc}
    ~	&\multicolumn{2}{c|}{AlP}&\multicolumn{2}{c|}{AlAs}&\multicolumn{2}{c|}{AlSb}&\multicolumn{2}{c|}{GaP}&\multicolumn{2}{c|}{GaAs}&\multicolumn{2}{c|}{GaSb}&	\multicolumn{2}{c|}{InP}&\multicolumn{2}{c|}{InAs}&	\multicolumn{2}{c}{InSb}\\ 
    
    \hline
    $\lambda_a$	&	0.0207	&(-0.06)&	0.1715	&( 0.00)&	0.3874	&( 0.01)	&	0.0307	&(-0.02)&	0.1813	&( 0.01)&	0.4504	&( 0.01)	&	0.0229	&( 0.00)&	0.175	&( 0.01)&	0.4518	&(-0.01)\\ 
    $\lambda_c$	&	0.0073	&( 0.00)&0.0072	&( 0.00)&	0.0079	&( 0.00)	&	0.041	&( 0.00)&	0.0408	&( 0.00)&	0.0433	&( 0.00)	&	0.1131	&(-0.01)&	0.1248	&( 0.00)&	0.1227	&( 0.00)\\ 
    \hline
    $E_{s,a}$	&	-6.6058	&(-0.19)&	-6.9122	&(-0.01)&	-6.0023	&(-0.07)&	-6.1425	&(-0.04)&	-6.0245	&(-0.02)&	-5.4982	&( 0.01)	&	-5.1873	&( 0.03)&	-5.7538	&( 0.06)&	-6.2602	&(-0.20)\\ 
    $E_{p,a}$	&	2.8082	&( 0.09)&	2.7705	&(-0.01)&	3.0269	&( 0.08)&	2.441	&( 0.10)&	3.6675	&( 0.00)&	3.224	&( 0.06)	&	3.3852	&(-0.02)&	3.3027	&( 0.03)&	3.7206	&( 0.01)\\ 
    $E_{d_{12},a}$	&	14.6773	&(-0.05)&	14.074	&(-0.14)&	12.0702	&(-0.11)&	15.3096	&(-0.13)&	12.6401	&( 0.04)&	11.1386	&(-0.01)	&	11.8032	&( 0.07)&	11.5632	&( 0.04)&	11.2501	&(-0.03)\\ 
    $E_{d_{15},a}$	&	11.6797	&( 0.16)&	10.4503	&( 0.13)&	10.8595	&(-0.01)&	11.0311	&( 0.17)&	13.692	&(-0.04)&	11.4367	&(-0.04)	&	12.5196	&( 0.02)&	12.4471	&(-0.03)&	10.6482	&( 0.03)\\ 
    $E_{s^*,a}$	&	22.4105	&(-0.10)&	19.8662	&(-0.06)&	16.4681	&(-0.05)&	19.3548	&( 0.02)&	19.131	&( 0.02)&	16.369	&(-0.02)	&	18.9953	&(-0.01)&	17.2566	&( 0.02)&	16.8295	&(-0.06)\\ 
    \hline																				
    $E_{s,c}$	&	0.7178	&(-0.01)&	0.1144	&(-0.01)&	0.2361	&( 0.00)&	-0.9678	&( 0.00)&	-0.3144	&( 0.00)&	-1.0009	&( 0.00)	&	0.3009	&( 0.00)&	0.1797	&( 0.00)&	0.0426	&(-0.02)\\ 
    $E_{p,c}$	&	6.5429	&(-0.07)&	5.5104	&( 0.00)&	5.8106	&(-0.03)&	6.2643	&(-0.07)&	6.9067	&(-0.07)&	6.0132	&(-0.04)	&	6.9977	&(-0.08)&	6.5234	&(-0.03)&	5.1306	&( 0.16)\\ 
    $E_{d_{12},c}$	&	11.8477	&( 0.15)&	10.4527	&( 0.13)&	10.6303	&( 0.01)&	12.1366	&( 0.10)&	12.0362	&( 0.09)&	10.2238	&( 0.07)	&	11.865	&( 0.07)&	10.5425	&( 0.12)&	9.1053	&( 0.16)\\ 
    $E_{d_{15},c}$	&	16.5362	&(-0.18)&	13.7099	&(-0.12)&	11.8107	&(-0.09)&	17.4598	&(-0.28)&	13.4595	&(-0.02)&	11.978	&(-0.08)	&	12.3531	&( 0.03)&	12.1772	&(-0.01)&	12.5455	&(-0.14)\\ 
    $E_{s^*,c}$	&	20.3264	&( 0.00)&	18.5132	&( 0.01)&	15.456	&( 0.01)&	18.5051	&( 0.06)&	19.2691	&( 0.01)&	16.0499	&( 0.00)	&	18.5897	&( 0.01)&	17.0919	&( 0.03)&	12.7396	&( 0.19)\\ 
    \hline
    $s s \sigma$	&	-1.8382	&(-0.06)&	-1.8042	&(-0.04)&	-1.748	&(-0.08)	&	-1.8177	&(-0.07)&	-1.6535	&(-0.02)&	-1.4552	&(-0.06)	&	-1.452	&(-0.04)&	-1.4811	&( 0.00)&	-1.4654	&(-0.30)\\ 
    $s_a p_c \sigma$	&	2.7009	&(-0.03)&	2.5436	&(-0.01)&	2.2294	&( 0.14)	&	2.6111	&( 0.07)&	2.4913	&( 0.00)&	2.4044	&( 0.06)	&	2.1679	&( 0.00)&	2.1689	&( 0.06)&	2.0664	&( 0.19)\\ 
    $s_a d_c \sigma$	&	-2.2456	&( 0.11)&	-2.1428	&( 0.16)&	-2.6982	&( 0.03)	&	-2.2849	&( 0.18)&	-2.6738	&( 0.02)&	-2.5768	&( 0.01)	&	-2.507	&( 0.02)&	-2.5093	&( 0.03)&	-2.1788	&( 0.15)\\ 
    $s_a s^*_c \sigma$	&	-1.4678	&(-0.03)&	-1.2659	&( 0.00)&	-1.2312	&(-0.02)	&	-1.6108	&( 0.02)&	-1.527	&( 0.02)&	-1.6649	&( 0.00)	&	-1.2652	&( 0.02)&	-1.2576	&(-0.03)&	-1.4979	&( 0.20)\\ 
    $s_c p_a \sigma$	&	2.877	&(-0.03)&	2.857	&(-0.04)&	2.8108	&( 0.04)	&	3.1278	&(-0.05)&	2.9333	&( 0.00)&	2.729	&(-0.01)	&	2.549	&( 0.04)&	2.4785	&( 0.12)&	2.3578	&( 0.13)\\ 
    $p p \sigma$	&	4.7281	&(-0.17)&	4.4956	&(-0.06)&	4.3744	&(-0.07)	&	4.5829	&(-0.09)&	4.7335	&(-0.07)&	4.5108	&(-0.01)	&	4.2026	&(-0.05)&	4.1402	&(-0.01)&	3.9329	&( 0.06)\\ 
    $p p \pi$	&	-1.4234	&(-0.09)&	-1.5647	&(-0.17)&	-1.6529	&(-0.08)	&	-1.6192	&(-0.13)&	-1.3242	&( 0.09)&	-1.7787	&(-0.06)	&	-1.1877	&( 0.07)&	-1.3302	&( 0.03)&	-1.412	&( 0.04)\\ 
    $p_a d_c \sigma$	&	-1.8547	&(-0.11)&	-2.014	&(-0.17)&	-2.240	&(-0.13)	&	-2.1733	&(-0.21)&	-1.8308	&(-0.03)&	-2.2774	&(-0.12)	&	-1.9619	&(-0.02)&	-2.1582	&(-0.02)&	-2.1305	&( 0.01)\\ 
    $p_a d_c \pi$	&	1.2727	&( 0.32)&	1.3352	&( 0.25)&	1.6579	&( 0.22)	&	1.3028	&( 0.30)&	1.7763	&( 0.00)&	1.7778	&( 0.10)	&	1.5683	&( 0.00)&	1.5602	&(-0.01)&	1.8582	&(-0.01)\\ 
    $s^*_c p_a \sigma$	&	2.2671	&( 0.03)&	2.1988	&( 0.00)&	1.8935	&( 0.00)	&	2.1737	&(-0.01)&	2.2339	&(-0.01)&	2.461	&( 0.00)	&	2.0919	&(-0.02)&	1.8675	&( 0.02)&	2.1124	&( 0.10)\\ 
    $s_c d_a \sigma$	&	-2.3231	&(-0.07)&	-2.5744	&(-0.08)&	-2.058	&(-0.03)	&	-2.6466	&(-0.14)&	-2.469	&(-0.02)&	-2.4401	&(-0.01)	&	-2.3094	&(-0.04)&	-2.4926	&(-0.02)&	-2.1118	&( 0.09)\\ 
    $p_c d_a \sigma$	&	-1.9477	&(-0.07)&	-1.9571	&(-0.11)&	-2.0745	&(-0.05)	&	-1.9942	&(-0.10)&	-1.7782	&( 0.01)&	-2.2698	&(-0.01)	&	-1.8735	&( 0.01)&	-1.9785	&( 0.04)&	-2.2119	&(-0.02)\\ 
    $p_c d_a \pi$	&	1.599	&( 0.27)&	1.491	&( 0.29)&	1.3758	&( 0.25)	&	1.577	&( 0.26)&	1.9289	&( 0.07)&	1.6354	&( 0.12)	&	1.7121	&( 0.04)&	1.6614	&( 0.03)&	1.6816	&( 0.09)\\ 
    $d d \sigma$	&	-1.3196	&( 0.02)&	-1.1657	&( 0.04)&	-1.1245	&( 0.01)	&	-1.1846	&( 0.03)&	-1.1431	&( 0.00)&	-1.2482	&( 0.00)	&	-1.2563	&(-0.01)&	-1.2026	&( 0.00)&	-1.2805	&( 0.02)\\ 
    $d d \pi$	&	2.5194	&(-0.06)&	2.1358	&( 0.02)&	2.0539	&( 0.03)	&	2.539	&(-0.12)&	2.3093	&(-0.05)&	2.2402	&(-0.02)	&	2.3329	&(-0.09)&	2.2768	&(-0.04)&	2.0377	&( 0.02)\\ 
    $d d \delta$	&	-1.6022	&( 0.13)&	-1.7413	&( 0.01)&	-1.5838	&( 0.08)	&	-2.0501	&(-0.02)&	-1.9845	&( 0.00)&	-1.6931	&( 0.03)	&	-1.7761	&(-0.05)&	-1.7433	&( 0.02)&	-1.3329	&( 0.06)\\ 
    $s^*_c d_a \sigma$	&	-0.712	&( 0.01)&	-0.7354	&( 0.01)&	-0.7758	&( 0.02)	&	-0.6476	&( 0.02)&	-0.6457	&( 0.00)&	-0.7981	&( 0.00)	&	-0.8095	&( 0.01)&	-0.83	&( 0.01)&	-0.794	&( 0.03)\\ 
    $s_c s^*_a \sigma$	&	-1.6871	&(-0.03)&	-1.4652	&( 0.09)&	-1.4931	&( 0.12)	&	-1.4465	&( 0.10)&	-2.0249	&(-0.02)&	-1.8242	&( 0.08)	&	-1.8984	&(-0.03)&	-1.9831	&( 0.07)&	-1.8296	&( 0.09)\\ 
    $s^*_a p_c \sigma$	&	1.9896	&( 0.04)&	2.1782	&(-0.03)&	2.4216	&( 0.02)	&	2.185	&( 0.09)&	2.1337	&( 0.02)&	3.0763	&(-0.02)	&	2.5094	&( 0.02)&	2.7347	&(-0.03)&	3.089	&(-0.13)\\ 
    $s^*_a d_c \sigma$	&	-0.7957	&(-0.02)&	-0.8239	&(-0.02)&	-0.7444	&(-0.02)	&	-0.6527	&(-0.02)&	-0.6942	&(-0.01)&	-0.8596	&( 0.00)	&	-0.7918	&( 0.00)&	-0.8583	&(-0.01)&	-0.7608	&(-0.03)\\ 
    $s^* s^* \sigma$	&	-3.7871	&(-0.04)&	-3.6196	&( 0.00)&	-3.2994	&( 0.00)	&	-3.59	&(-0.01)&	-3.8005	&(-0.03)&	-3.2367	&( 0.00)	&	-3.7707	&(-0.02)&	-4.0354	&(-0.05)&	-4.0693	&(-0.26)\\ 
    \hline
    $\Delta E_{n.i.}$& -0.0130&-&0.0130&-&0.0135&-&-0.0057&-&-0.0086&-&-0.0025&-& -0.0007&-& 0.0015&-&0.0033&-\\
    \end{tabular}
	\caption[WZ III-V SK-Parameters and ZB Comparison]{\small WZ III-V SK-parameters in eV. Values in parenthesis are the fractional difference between our WZ parameters and the ZB parameters from Ref. \onlinecite{Jancu.prb.1998}. The onsite energies have been shifted to make the top of the valence band have zero energy ($E(\Gamma_{9v})=0$) for the ideal WZ crystal (Table \ref{tab:lat_const_table}). $\Delta E_{n.i.}=E'(\Gamma_{9v})-E(\Gamma_{9v})$ is the shift in the valence band maximum when going from ideal to non-ideal WZ. To obtain the parameters for the non-ideal structure, subtract $\Delta E_{n.i}$ from the ideal onsite energies.}\label{tab:III-V-series-2H_params}
\end{sidewaystable}
}

\onecolumngrid
~

\begin{table}[!ht]
    \centering
    \small
    \begin{tabular}{c|c|c|c|c  ||c|c|c|c|c  ||c|c|c|c|c}
        \multicolumn{5}{c||}{C}&\multicolumn{5}{c||}{Si}&\multicolumn{5}{c}{Ge}\\ 
        \hline
        &&\multicolumn{2}{c|}{TB}&&&&\multicolumn{2}{c|}{TB}&&&&\multicolumn{2}{c|}{TB}&\\
        &EPP$^a$&Ideal&Non-Ideal&Exp&&EPP$^a$&Ideal&Non-Ideal&Exp&&EPP$^a$&Ideal&Non-Ideal&Exp\\
        \hline
        $\Gamma_{7v}^{+}$   &-1.6965&-1.2882&-1.1583&-  &$\Gamma_{7v}^{+}$   &-0.3496&-0.9726&-0.9022&- &$\Gamma_{7v}^{+}$   &-0.4896&-0.4835&-0.4648&-\\ 
        $\Gamma_{7v}^{+}$   &-0.003&-0.0052&-0.0052&-   &$\Gamma_{7v}^{+}$   &-0.0279&-0.031&-0.0341&-  &$\Gamma_{7v}^{+}$   &-0.1293&-0.1321&-0.1256&-\\ 
        $\Gamma_{9v}^{+}$   &0.000&0.000&0.000&-        &$\Gamma_{9v}^{+}$   &0.000&0.000&0.000&-       &$\Gamma_{9v}^{+}$   &0.000&0.000&0.000&-\\ 
        $\Gamma_{7c}^{-}$   &5.7365&5.7665&5.9015&-     &$\Gamma_{8c}^{-}$   &1.4814&1.6703&1.8333&-    &$\Gamma_{8c}^{-}$   &0.3103&0.3102&0.3666&-\\
        $\Gamma_{9c}^{-}$   &5.741&5.7712&5.9062&-      &$\Gamma_{7c}^{-}$   &2.6962&2.442&2.4875&-     &$\Gamma_{7c}^{-}$   &0.7659&0.7965&0.7945&-\\ 
        \hline
        $K_{c}$             &4.7676&7.4955&7.5558&-           &$K_{c}$             &1.2191&2.2423&2.3914&-           &$K_{c}$             &1.5267&0.9908&1.0523&-\\ 
        $M_{c}$             &5.2688&5.2907&5.288&-            &$M_{c}$             &0.7957&0.7962&0.8309&-           &$M_{c}$             &0.6828&0.682&0.7438&-\\ 
        $A_{7c}$            &7.2555&4.7961&5.0482&-           &$A_{8c}$            &2.3662&2.2644&2.2645&-           &$A_{8c}$            &1.0951&3.5758&3.6067&-\\ 
        $H_{c}$             &7.5599&7.3609&7.3229&-           &$H_{c}$             &2.3229&1.5048&1.4877&-           &$H_{c}$             &1.7489&1.0716&1.0449&-\\ 
        $L_{c}$             &5.5528&6.015&5.9343&-            &$L_{c}$             &1.2958&1.3375&1.3834&-           &$L_{c}$             &0.7849&0.7819&0.8064&-\\ 
        \hline
        $m_l(\Gamma_{9v}^{+})$  &-0.3418&-0.5463&-0.5639   &-&$m_l(\Gamma_{9v}^{+})$&-0.5637&-0.5861&-0.6059   &-&$m_l(\Gamma_{9v}^{+})$&-0.6035&-0.4073&-0.4191&-\\ 
        $m_l(\Gamma_{7c}^{-})$  &0.7860&0.9972&1.0545     &-&$m_l(\Gamma_{8c}^{-})$&1.0483&1.5706&1.607      & -&$m_l(\Gamma_{8c}^{-})$&1.0563&1.1443&1.1176&-\\ 
        $m_t(\Gamma_{9v}^{+})$  &-0.3226&-0.3055&-0.3049   &- &$m_t(\Gamma_{9v}^{+})$&-0.2128&-0.1817&-0.2179   &-&$m_t(\Gamma_{9v}^{+})$&-0.0672&-0.0559&-0.0578&-\\ 
        $m_t(\Gamma_{7c}^{-})$  &0.3785&0.4373&0.4318     &-&$m_t(\Gamma_{8c}^{-})$&0.1224&0.1615&0.163      &-&$m_t(\Gamma_{8c}^{-})$&0.0852&0.1086&0.1067&-
    \end{tabular}
    \begin{tabular}{l}
        $^a$Reference \onlinecite{De.IOP.2014}~~~~~~~~~~~~~~~~~~~~~~~~~~~~~~~~~~~~~~~~~~~~~~~~~~~~~~~~~~~~~~~~~~~~~~~~~~~~~~~~~~~~~~~~~~~~~~~~~~~~~~~~~~~~~~~~~~~~~~~~~~~~~~~~~~~~~~~~~~~~~~~~~~~~~~~~~~~~~~~~~~~~\\
    \end{tabular}
    \caption{Energies ($\rm{eV}$) and effective masses ($\rm{m_e}$) for ideal and non-ideal (Table \ref{tab:lat_const_table}) LDL calculated using the present ETB parameterization, and EPP data from Ref. \onlinecite{De.IOP.2014}. Experimental values are shown when available and marked with a `-' otherwise. Energy values are referenced to the top of the valence band ($\Gamma_{9v}$).
}
    \label{tab:my_label}
\end{table}

%2H Al-series comparison
\begin{table}[ht!]
    \centering
    \small
    \begin{tabular}{c||c|cc|c||c|cc|c||c|cc|c}						
        &\multicolumn{4}{c||}{AlP}&  \multicolumn{4}{c||}{AlAs}&  \multicolumn{4}{c}{AlSb}\\
        \hline
                                 &$PP^a$     &\multicolumn{2}{c|}{TB}&Exp.&$PP^a$  & \multicolumn{2}{c|}{TB}&Exp.&$PP^a$  & \multicolumn{2}{c|}{TB}&Exp.\\
                                 &         & Ideal&Non-Ideal&&& Ideal&Non-Ideal&&& Ideal&Non-Ideal&\\
        \hline 
        $E_g$           &2.969&2.955&2.997     &-&1.971&1.966&1.951      &-&1.891&1.8907&1.8567&-\\
        \hline
        $\Gamma_{7v'}$  &-0.435&-0.148&-0.118   &-   &-0.518&-0.467&-0.457  &-&-0.802&-0.900&-0.891&-\\ 
        $\Gamma_{7v}$   &-0.044&-0.028&-0.027   &-&-0.139&-0.134&-0.128  &-&-0.156&-0.156&-0.145&-\\ 
        $\Gamma_{9v}$   &0.000&0.000&0.000           &-&0.000 &0.000&0.000          &-&0.000&0.000&0.000&-\\ 
        $\Gamma_{8c}$   &2.969&2.955&2.997     &-&1.971&1.966&1.951      &-&1.891&1.891&1.857&-\\ 
        $\Gamma_{7c}$   &3.775&3.500&3.526     &-&3.153&3.007&2.916     &-&2.418&2.553&2.455&-\\ 
        \hline
        $K_{c}$         &3.823&3.523&3.568     &-&2.932&2.605&2.594    &-&2.355&2.414&2.385&-\\ 
        $M_{c}$         &3.157&3.141&3.186     &-&2.146&2.140&2.185      &-&1.929&1.930&1.972&-\\ 
        $A_{c}$         &3.684&5.205&5.205     &-&2.745&4.135&4.162     &-&2.478&3.676&3.722&-\\ 
        $H_{c}$         &4.446&3.538&3.527     &-&3.478&2.528&2.494    &-&2.950&2.290&2.270&-\\ 
        $L_{c}$         &3.253&3.275&3.296      &-&2.354&2.333&2.325    &-&2.060&2.108&2.099&-\\ 
        \hline
        $m_l(\Gamma_{9v})$&-0.972&-0.599&-0.605 &-&-0.933&-0.508&-0.519  &-&-0.959&-0.523&-0.534&-\\ 
        $m_l(\Gamma_{8c})$& 1.187&1.397&1.379    &-&1.081&1.198&1.168     &-&1.160&1.210&1.173&-\\ 
        $m_t(\Gamma_{9v})$&-0.248&-0.217&-0.237 &-&-0.216&-0.174&-0.176  &-&-0.211&-0.155&-0.155&-\\ 
        $m_t(\Gamma_{8c})$&0.170&0.216&0.215    &-&0.142&0.1703&0.169     &-&0.157&0.148&0.146&-\\
        \multicolumn{13}{c}{~}\\
        &\multicolumn{4}{c||}{GaP}&  \multicolumn{4}{c||}{GaAs}&  \multicolumn{4}{c}{GaSb}\\
        \hline
                                 &$PP^a$     &\multicolumn{2}{c|}{TB}&Exp.&$PP^a$  & \multicolumn{2}{c|}{TB}&Exp.&$PP^a$  & \multicolumn{2}{c|}{TB}&Exp.\\
                                 &         & Ideal&Non-Ideal&&& Ideal&Non-Ideal&&& Ideal&Non-Ideal&\\
        \hline 
        $E_g$                       &~2.251 &~2.051&~2.096&~2.108$^a$&~1.503&~1.503&~1.521&~1.522$^b$&&~0.512&~0.554&-\\
        \hline
        \hline
        $\Gamma_{7v'}$  &-0.373&-0.186&-0.178&-  &-0.475&-0.497&-0.486&-   &-0.874&-1.004&-1.004&-\\ 
        $\Gamma_{7v}$   &-0.050&-0.045&-0.044&-  &-0.120&-0.132&-0.127&-0.104$^c$  &-0.142&-0.149&-0.149&-\\ 
        $\Gamma_{9v}$   &~0.000&~0.000   &~0.000   &-    &~0.000 &~0.000&~0.000       &-   &~0.000&~0.000&~0.000&-\\ 
        $\Gamma_{8c}$   &~2.251 &~2.051&~2.096&~2.108$^a$    &~1.503&~2.170&~2.216&~1.58$^c$    &~0.509 &~0.512&~0.554&-\\ 
        $\Gamma_{7c}$   &~2.877&~2.866&~2.895&-    &~1.588&~1.503&~1.521&~1.522$^b$,~1.52$^c$     &~0.851&~0.877&~0.904&-\\ 
        \hline
        $K_{c}$         &~3.054&~2.637&~2.685&-    &~3.282&~2.675&~2.724&-    &~1.442&~1.121&~1.167&-\\ 
        $M_{c}$         &~2.402&~2.376&~2.415&-    &~2.167&~2.144&~2.201&-    &~0.785&~0.788&~0.820&-\\ 
        $A_{c}$         &~2.986&~4.586&~4.569&-    &~2.257&~4.300&~4.291&-    &~1.237&~3.083&~3.079&-\\ 
        $H_{c}$         &~3.653&~2.812&~2.787&-    &~3.318&~2.755&~2.725&-    &~1.890&~1.164&~1.134&-\\ 
        $L_{c}$         &~2.587&~2.575&~2.588&-    &~2.136&~2.209&~2.229&-    &~0.858&~0.931&~0.945&-\\ 
        \hline
        \hline
        $m_l(\Gamma_{9v})$&-0.941&-0.558&-0.561&-    &-1.026&-0.733&-0.746&-  &-0.833&-0.502&-0.511&-\\ 
        $m_l(\Gamma_{8c})$&~1.162 &~1.431&~1.415&-      &-     &-     &-      &-   &~0.983&~1.090&~1.075&-\\ 
        $m_l(\Gamma_{7c})$&-     &-     &-     &-      &~0.090 &~0.081&~0.079 & - &-    &-     &-&-\\ 
        \hline
        $m_t(\Gamma_{9v})$&-0.205&-0.178&-0.191&-    &-0.134&-0.113&-0.118 &-   &-0.087&-0.068&-0.071&-\\ 
        $m_t(\Gamma_{8c})$&~0.143 &~0.170 &~0.169 &-    &-     &-      &-       &-   & ~0.096&~0.100 &~0.098&-\\
        $m_t(\Gamma_{7c})$&-     &-      &-      &-   &~0.082 &~0.073  &~0.075  &  -&-     &-      &-&-\\
        \multicolumn{13}{l}{$^a$ Reference \onlinecite{Kriegner.PRB.13}}\\
        \multicolumn{13}{l}{$^b$ Reference \onlinecite{DeLuca.NanoLetters.17}}\\
        \multicolumn{13}{l}{$^c$ Reference \onlinecite{Vainorius.Nanoscale.18}}\\
        &\multicolumn{4}{c||}{InP}&  \multicolumn{4}{c||}{InAs}&  \multicolumn{4}{c}{InSb}\\
        \hline
                                 &$PP^a$     &\multicolumn{2}{c|}{TB}&Exp.&$PP^a$  & \multicolumn{2}{c|}{TB}&Exp.&$PP^a$  & \multicolumn{2}{c|}{TB}&Exp.\\
                                 &         & Ideal&Non-Ideal&&& Ideal&Non-Ideal&&& Ideal&Non-Ideal&\\
        \hline 
        $E_g$                       &1.474   &1.487 &1.491 &1.492$^d$,1.508$^f$  &0.481   &0.480 &0.480&0.482 $^e$    &0.287 &0.288 &0.286&\\
        \hline
        \hline
        $\Gamma_{7v'}$  &-0.348  &-0.338&-0.339&-  &-0.469  &-0.490&-0.494&-0.433$^e$    &-0.847   &-0.884 &-0.888&-\\ 
        $\Gamma_{7v}$   &-0.063  &-0.061&-0.062&-.036$^d$   &-0.105  &-0.111&-0.114&-0.068$^e$    &-0.098   &-0.101&-0.108&-\\ 
        $\Gamma_{9v}$   &0.000   &0.000    &0.000    &-   &0.000   &0.000    &0.000    &-    &0.000    &0.000    &0.000    &-\\ 
        $\Gamma_{7c}$   &1.474   &1.487 &1.491 &-   &0.481   &0.480 &0.480 &0.482$^e$    &0.287    &0.288 &0.286 &-\\ 
        $\Gamma_{8c}$   &1.712   &2.061  &2.080 &-   &1.222   &1.322 &1.339 &1.065$^e$    &1.116    &0.621 &0.634  &-\\ 
        \hline
        $K_{c}$         &2.761  &2.528 &2.548 &-  &2.174   &1.686 &1.703&-   &0.766   &0.876 &0.886&-\\ 
        $M_{c}$         &2.059  &2.054 &2.078 &-  &1.554   &1.544 &1.569&-   &0.910   &0.911 &0.928&-\\ 
        $A_{c}$         &2.347  &3.892 &3.886 &-  &1.659   &3.585 &3.581&-   &1.336   &2.984 &2.971&-\\ 
        $H_{c}$         &3.186  &2.574 &2.553  &-  &2.742   &2.000 &1.977&-   &2.049   &1.133 &1.113&-\\ 
        $L_{c}$         &2.120  &2.165 &2.171 &-  &1.661   &1.638 &1.644&-   &1.066   &0.908 &0.910&-\\ 
        \hline
        \hline
        $m_l(\Gamma_{9v})$&-1.273&  -0.864 &-0.874&-  &-1.700 &-0.744&-0.751&-  &-2.060 &-0.540&-0.547&-\\ 
        $m_l(\Gamma_{7c})$& 0.105&  0.097  &0.096 &-  & 0.060 &0.047 &0.047 &-  &0.051  &0.047 &0.048 &-\\ 
        \hline
        $m_t(\Gamma_{9v})$&-0.158&  -0.141 &-0.147&-  &-0.084 &-0.057&-0.059&-  &- 0.066&-0.042&-0.044&-\\ 
        $m_t(\Gamma_{7c})$&0.088&   0.085  &0.086 &-  & 0.042 &0.037 &0.037 &-  & 0.035 &0.033  &0.033 &-\\
        \multicolumn{13}{l}{$^d$ Reference \onlinecite{Chauvin.APL.18}}\\
        \multicolumn{13}{l}{$^e$ Reference \onlinecite{Pournia.NanoRes.2020}}\\
        \multicolumn{13}{l}{$^f$ Reference \onlinecite{Perera.NanoLetters.13}}
    \end{tabular}
    \caption[III-V-comparison-tables]{\small Energies ($\rm{eV}$) and effective masses ($\rm{m_e}$) for ideal and non-ideal (Table \ref{tab:lat_const_table}) WZ calculated using the present ETB parameterization, and EPP data from Ref. \onlinecite{De.IOP.2014}. Experimental values are shown when available and marked with a `-' otherwise. Energy values are referenced to the top of the valence band ($\Gamma_{9v}$)}.\label{tab:III-V-comparison-tables}
\end{table}

\onecolumngrid
{\setlength{\tabcolsep}{0.25em}
\begin{table}[ht!]
    \centering
    \small
    \begin{tabular}{c|c|c|c||  c|c|c|c||  c|c|c|c}
         \multicolumn{4}{c||}{C}&\multicolumn{4}{c||}{Si}&\multicolumn{4}{c}{Ge}\\ 
         \hline
         ~&Weight&Target&Result&~&Weight&Target&Result&~&Weight&Target&Result\\ 
        \hline
        $\Delta_{AB} [\rm{eV}]$        &10&0.003  &0.005  &$\Delta_{AB}$        &10&0.028&0.031               &$\Delta_{AB}$&50&0.129&0.132\\
        $K_{9c}- M_{c}$     &50&-0.501 &-0.495 &$M_c$                &1000&0.796&0.796             &$\Gamma_{8c}^{-}$&1000&0.310&0.310\\
        $K_{9c}\shortminus\Gamma_{c}$&500&-0.969 &-0.97 &$M_{c}\shortminus\Gamma_{c}$ &100&-0.686&-0.874            &$LM_{c}\shortminus M_{c}$&50&-0.064&-0.064\\ 
        $K_{9c}$             &1000&4.768 &4.796 &$LM_{c}\shortminus M_{c}$     &100&0.382&0.28               &$LM_{c}\shortminus\Gamma_{c}$&500&0.309&0.308\\
        $\Gamma_{7c}^{\shortminus}$    &10&5.737  &5.766  &$M_{c}\shortminus K_{c}$      &10&-0.423&-1.468             &~&~&~&~\\
        $\Gamma_{7c'}^{\shortminus}$   &10&8.825  &6.353  &$\Gamma_{7c}^{\shortminus}\shortminus\Gamma_{8c}^{\shortminus}$&100&1.215&0.772&~&~&~&~\\
                             ~&~&~      &~     &$H_{c}$-$K_{c}$     &10&1.104      &-0.76          &~&~&~&~\\
        \hline
        $m_l(\Gamma_{7c})$  &10&0.786  &0.997  &-&-   &-&-&                                         $m_l(\Gamma_{7c})$&10&1.056&1.144\\
        $m_l(\Gamma_{7c'})$   &10&0.25   &0.285  &-&-   &-&-&                                       $m_l(\Gamma_{8c})$&10&0.052&0.049\\
    \end{tabular}
    \caption{Group-IV Target Comparison.}\label{tab:IV_target_values}
    \begin{tabular}{c||c|cc|cc|cc}	
        &&\multicolumn{2}{c|}{AlP}&\multicolumn{2}{c|}{AlAs}&\multicolumn{2}{c}{AlSb}\\
        \hline
        &Weight&Target&Result&Target&Result&Target&Result\\
        \hline
        $\Delta_{AB}$       &10     &0.044  &0.027  &0.139  &0.134  &0.156  &0.156\\
        $\Gamma_{8c}$       &1000   &2.969  &2.955  &1.971  &1.966  &1.891  &1.891\\
        $M_{c}$-$\Gamma_{c}$&500    &0.187  &0.186  &0.175  &0.174  &0.04   &0.039\\
        $LM_{c}$-$M_{c}$    &10     &0.039  &0.035  &0.161  &0.078  &0.048  &0.046\\
        \hline
        $m_l(\Gamma_{7c})$  &10     &1.187  &1.397  &1.081  &1.198  &1.16   &1.21\\
        $m_l(\Gamma_{8c})$  &10     &0.182  &0.182  &0.18   &0.159  &0.163  &0.152
    \end{tabular}
    \caption{\small Al-series Target Comparison.}\label{tab:Al_target_values}
    \begin{tabular}{c||c|cc|cc|cc}	
        &&\multicolumn{2}{c|}{GaP}&\multicolumn{2}{c|}{GaAs}&\multicolumn{2}{c}{GaSb}\\
        \hline
        &Weight&Target&Result&Target&Result&Target&Result\\
        \hline
        $\Delta_{AB}$           &10     &0.05   &0.044  &0.12   &0.126  &0.142  &0.149\\
        $\Gamma_{8c}$           &1000   &2.108$^a$  &2.098  &-      &-      &0.509  &0.512\\
        $\Gamma_{7c}$           &1000   &-      &-      &1.522$^b$  &1.522  &-      &-\\ 
        $LM_{c}$-$\Gamma_{c}$   &500    &-      &-      &0.532  &0.531  &-0.044 &-0.046\\
        $M_{c}$-$\Gamma_{c}$    &500    &0.298  &0.294  &-      &-      &-      &-\\
        $LM_{c}$-$M_{c}$        &10     &0.074  &0.053  &-0.109 &-0.112 &0.277  &0.276\\
        \hline 
        $m_l(\Gamma_{7c})$      &10     &1.162  &1.415  &1.05   &1.284  &0.983  &1.09\\
        $m_l(\Gamma_{8c})$      &10     &0.153  &0.129  &0.09   &0.079  &0.064  &0.056\\
        \multicolumn{1}{l}{$^a$ Reference \onlinecite{Kriegner.PRB.13}}\\
        \multicolumn{1}{l}{$^b$ Reference \onlinecite{Vainorius.Nanoscale.18}}\\
    \end{tabular}
    \caption{\small Ga-series Target Comparison. }\label{tab:Ga_target_values}
    \begin{tabular}{c||c|cc|cc|cc}	
        &&\multicolumn{2}{c|}{InP}&\multicolumn{2}{c|}{InAs}&\multicolumn{2}{c}{InSb}\\
        \hline
        &Weight&Target&Result&Target&Result&Target&Result\\
        \hline
        $\Delta_{AB}$           &10     &0.063  &0.061  &0.105  &0.113  &0.098  &0.101  \\
        $\Gamma_{7c}$           &1000   &1.492$^a$  &1.493  &0.481$^b$  &0.482  &0.287  &0.288  \\
        $LM_{c}$-$\Gamma_{c}$   &500    &0.501  &0.5    &0.969  &0.968  &-0.165 &-0.165 \\
        $LM_{c}$-$M_{c}$        &10     &-0.067 &-0.068 &-0.103 &-0.104 &0.458  &0.457  \\
        \hline
        $m_l(\Gamma_{7c})$      &10     &0.105  &0.095  &0.06   &0.046  &0.051  &0.047  \\
        $m_l(\Gamma_{8c})$      &10     &1.094  &1.278  &1.276  &1.392  &1.781  &1.791  \\
        \multicolumn{1}{l}{$^a$ Reference \onlinecite{Chauvin.APL.18}}\\
        \multicolumn{1}{l}{$^b$ Reference \onlinecite{Pournia.NanoRes.2020}}\\
    \end{tabular}
    \caption{\small In-series Target Comparison.}\label{tab:In_target_values}
    {Targets in units of $\rm{eV}$ and $\rm{m_e}$, weights used during fitting and resulting fits. $LM_c$ refers to the valley along the L-M path. A minimal set of targets was used in order to keep the WZ corrections to the ZB parameters small. Carbon and Silicon required additional constraints to avoid unphysical band structures, such as extrema with wrong symmetries. Cells marked with a `-' were not used as a targets for that material. Unless otherwise noted, target values are EPP calculations from Refs. \onlinecite{De.PRB.2010,De.IOP.2014}. }
\end{table}
}
~
\twocolumngrid
~

\onecolumngrid
~
{\setlength{\tabcolsep}{0.5em}
\begin{table}[ht!]
    \centering
    \begin{tabular}{c|ccc||c|ccc}
    \multicolumn{4}{c||}{ZB}&\multicolumn{4}{c}{WZ}\\
        ~&AlP&	        AlAs&	        AlSb&~&AlP&	        AlAs&	        AlSb\\
        \hline
        $\frac{d}{dp}E(\Gamma_{8v}-\Gamma_{6c})$&9.27&	9.39&	12.21&$\frac{d}{dp}E(\Gamma_{9v}-\Gamma_{8c})$&8.66&	8.61&	11.14\\
        $\frac{d}{dp}E(\Gamma_{8v}-L_{6c})$&4.69&	4.67&	6.77&$\frac{d}{dp}E(\Gamma_{9v}-\Gamma_{7c})$&3.90&	3.81&	5.53\\
        \multicolumn{4}{c||}{}\\
        ~&GaP&	        GaAs&	        GaSb&~&GaP&	        GaAs&	        GaSb\\
        \hline
        $\frac{d}{dp}E(\Gamma_{8v}-\Gamma_{6c})$&10.49&	11.62&	14.03&$\frac{d}{dp}E(\Gamma_{9v}-\Gamma_{8c})$&9.77&  11.34&	13.95\\
        $\frac{d}{dp}E(\Gamma_{8v}-L_{6c})$&4.69&	4.73&	5.10&$\frac{d}{dp}E(\Gamma_{9v}-\Gamma_{7c})$&3.70&  4.73&	4.71\\
        \multicolumn{4}{c||}{}\\
        ~&InP&	        InAs&	        InSb&~&InP&	        InAs&	        InSb\\
        \hline
        $\frac{d}{dp}E(\Gamma_{8v}-\Gamma_{6c})$&9.11&	10.55&	13.97&$\frac{d}{dp}E(\Gamma_{9v}-\Gamma_{8c})$&9.04&	9.22&	9.92\\
        $\frac{d}{dp}E(\Gamma_{8v}-L_{6c})$&3.74&	4.40&	4.88&$\frac{d}{dp}E(\Gamma_{9v}-\Gamma_{7c})$&3.57&	3.61&	4.48\\
    \end{tabular}
    \caption{Pressure coefficients in $\rm{meV/kBar}$ between the top of the valence band ($\Gamma_{8v}$) and two lowest conduction band valley states ($\Gamma_{6c},L_{6c}$) in ZB, and the top of the valence band ($\Gamma_{8v}$) and the lowest two conduction band states at $\Gamma$ ($\Gamma_{7c},\Gamma_{8c}$) in WZ. The pressure coefficients for the ZB crystal phase were calculated using the ETB and radial scaling parameters from Ref. \cite{Jancu.prb.1998}. The pressure coefficients for the WZ crystal phase were calculated using ETB parameters from the present work and radial scaling parameters from Ref. \cite{Jancu.prb.1998}. See Fig.\ref{fig:ZB-WZ_bandstructure_comp} for correspondence between ZB and WZ bands.}\label{table:def_pot_comp}
\end{table}
}
\twocolumngrid

\clearpage
%\bibliography{citations}
\providecommand{\noopsort}[1]{}\providecommand{\singleletter}[1]{#1}

\end{document}